\documentclass[11pt]{article}

\newsavebox{\foobox}
\newcommand{\slantbox}[2][0]{\mbox{%
        \sbox{\foobox}{#2}%
        \hskip\wd\foobox
        \pdfsave
        \pdfsetmatrix{1 0 #1 1}%
        \llap{\usebox{\foobox}}%
        \pdfrestore
}}
\newcommand\unslant[2][-.25]{\slantbox[#1]{$#2$}}

\newcommand{\mpi}{\text{\unslant[-.18]\pi}}
\newcommand{\mdelta}{\text{\unslant[-.18]\delta}}
\newcommand{\mepsilon}{\text{\unslant[-.15]\epsilon}}
\newcommand{\mgamma}{\text{\unslant[-.15]\gamma}}
\newcommand{\malpha}{\text{\unslant[-.18]\alpha}}

\newcommand{\mbeta}{\text{\unslant[-.25]\beta}}

\usepackage[left=2cm, right=2cm, top=2.5cm, bottom=2.5cm]{geometry}
\usepackage[x11names]{xcolor}
\usepackage{fancyhdr, amssymb, cancel, amsmath, graphicx, pgfplots, tikz}
\usepackage{isomath}

\usetikzlibrary{shadows}

\newcommand{\stylecolor}{blue}

\usepackage[labelfont={bf,sf, color=\stylecolor}, margin={1.5cm,0cm}]{caption}

\usepackage[colorlinks=true, urlcolor=\stylecolor!70!white, linkcolor=\stylecolor, citecolor=\stylecolor!70!white, hyperindex=true, linktocpage=true]{hyperref}

\usepackage{amsthm}
\newtheoremstyle{theor}{10pt}{10pt}{}{16pt}{\sffamily \bfseries \color{green!50!black}}{:}{.5em}{}
\theoremstyle{theor}

\usepackage[explicit]{titlesec}

\newcommand*\sectionlabel{}
\titleformat{\section}
  {\gdef\sectionlabel{}
   \Large\bfseries\scshape}
  {\gdef\sectionlabel{\thesection }}{0pt}
  {\begin{tikzpicture}[remember picture,overlay]
       \end{tikzpicture}
  }
\titlespacing*{\section}{0pt}{0pt}{0pt}

\newcommand*\subsectionlabel{}
\titleformat{\subsection}
  {\gdef\subsectionlabel{}
   \large\bfseries\scshape}
  {\gdef\subsectionlabel{\thesubsection  }}{0pt}
  {\begin{tikzpicture}[remember picture,overlay]
    	\draw (-0.15, 0.02) node[right] {\color{\stylecolor} \textsf{\subsectionlabel}};
	\draw (1.25, 0) node[right] {\color{\stylecolor} \textsf{#1}};
	\fill[color=\stylecolor] (1,-0.25) rectangle (1.1, 0.25);
       \end{tikzpicture}
  }
\titlespacing*{\subsection}{0pt}{10pt}{10pt}

\newcommand*\subsubsectionlabel{}
\titleformat{\subsubsection}
  {\gdef\subsubsectionlabel{}
   \bfseries\scshape}
  {\gdef\subsubsectionlabel{\thesubsubsection.\ \  }}{0pt}
  {\begin{tikzpicture}[remember picture,overlay]
    	\draw (-0.15, 0) node[right] {\color{\stylecolor} \textsf{\subsubsectionlabel#1}};
       \end{tikzpicture}
  }
\titlespacing*{\subsubsection}{0pt}{7pt}{7pt}

\pgfplotsset{every axis legend/.append style={at={(1.02,1)},anchor=north west}}

\begin{document}

\allowdisplaybreaks

\pagestyle{fancy}
\renewcommand{\headrulewidth}{0pt}
\fancyhead{}

\fancyfoot{}
\fancyfoot[C] {\textsf{\textbf{\thepage}}}

\begin{equation*}
\begin{tikzpicture}
\draw (\textwidth, 0) node[text width = \textwidth, right] {\color{white} easter egg};
\end{tikzpicture}
\end{equation*}

\begin{equation*}
\begin{tikzpicture}
\draw (0.5\textwidth, -3) node[text width = \textwidth] {\huge  \textsf{\textbf{Emergent entropy production and hydrodynamics in \\ \vspace{0.07in}  quantum many-body systems  }} };
\end{tikzpicture}
\end{equation*}
\begin{equation*}
\begin{tikzpicture}
\draw (0.5\textwidth, 0.1) node[text width=\textwidth] {\large \color{black} \textsf{Tom Banks}$^{\color{\stylecolor} \mathsf{a}}$  \textsf{and Andrew Lucas}$^{\color{\stylecolor} \mathsf{b}}$};
\draw (0.5\textwidth, -1) node[text width=\textwidth] {$^{\color{\stylecolor} \mathsf{b}}$ \small{\textsf{Department of Physics, Stanford University, Stanford, CA 94305, USA}}};
\draw (0.5\textwidth, -0.5) node[text width=\textwidth] {$^{\color{\stylecolor} \mathsf{a}}$ {\small \textsf{Department of Physics and NHETC, Rutgers University, Piscataway, NJ 08854, USA}}};
\end{tikzpicture}
\end{equation*}
\begin{equation*}
\begin{tikzpicture}
\draw (0, -13.1) node[right, text width=0.5\paperwidth] {\texttt{tibanks@ucsc.edu, ajlucas@stanford.edu}};
\draw (\textwidth, -13.1) node[left] {\textsf{\today}};
\end{tikzpicture}
\end{equation*}
\begin{equation*}
\begin{tikzpicture}
\draw[very thick, color=\stylecolor] (0.0\textwidth, -5.75) -- (0.99\textwidth, -5.75);
\draw (0.12\textwidth, -6.25) node[left] {\color{\stylecolor}  \textsf{\textbf{Abstract:}}};
\draw (0.53\textwidth, -6) node[below, text width=0.8\textwidth, text justified] {\small We study dynamics of a locally conserved energy in ergodic, local many-body quantum systems on a lattice with no additional symmetry.  The resulting dynamics is well approximated by a coarse grained, classical linear functional diffusion equation for the probability of all spatial configurations of energy.   This is equivalent to nonlinear stochastic hydrodynamics, describing the diffusion of energy in physical spacetime.  We find the absence of non-hydrodynamic slow degrees of freedom, a nonlinear fluctuation-dissipation theorem, and the emergence of a (weakly interacting) kinetic theory for hydrodynamic modes near thermal equilibrium.   The observable part of the microscopic entropy obeys the local second law of thermodynamics, and quantitatively agrees with the phenomenological predictions of hydrodynamics.   Our approach naturally generalizes to ergodic  systems with additional symmetries, may lead to numerical algorithms to calculate diffusion constants for lattice models, and implies sufficiency conditions for a rigorous derivation of hydrodynamics in  quantum systems.};
\end{tikzpicture}
\end{equation*}

\tableofcontents

\titleformat{\section}
  {\gdef\sectionlabel{}
   \Large\bfseries\scshape}
  {\gdef\sectionlabel{\thesection }}{0pt}
  {\begin{tikzpicture}[remember picture,overlay]
	\draw (1, 0) node[right] {\color{\stylecolor} \textsf{#1}};
	\fill[color=\stylecolor] (0,-0.35) rectangle (0.7, 0.35);
	\draw (0.35, 0) node {\color{white} \textsf{\sectionlabel}};
       \end{tikzpicture}
  }
\titlespacing*{\section}{0pt}{15pt}{15pt}

\begin{equation*}
\begin{tikzpicture}
\draw[very thick, color=\stylecolor] (0.0\textwidth, -5.75) -- (0.99\textwidth, -5.75);
\end{tikzpicture}
\end{equation*}

\section{Introduction}
Hydrodynamics is a universal description of the late time dynamics of a thermalizing system, regardless of whether the microscopic dynamics are classical or quantum.   From a phenomenological perspective, it is well understood how to write down the hydrodynamic equations \cite{landau, kadanoff}: in the simplest cases, they are simply the conservation laws for conserved quantities in the system.    Such conservation laws follow from a quantum mechanical Ward identity and are exact.   However, these equations are also highly underdetermined.   To obtain solvable equations of motion, one must relate the fluxes of conserved quantities to their densities.   This is done order by order in a derivative expansion, and the regime of validity of hydrodynamics describes long wavelength fluctuations relative to local thermalization length scales (``mean free paths").     Two final phenomenological requirements are then imposed.   Firstly, hydrodynamics must be compatible with the local second law of thermodynamics -- i.e., the theory is dissipative.   Secondly, the fluxes of conserved quantities must have stochastic contributions, consistent with the fluctuation dissipation relation.   We will quantify all of these points later.   For now, we simply stress that the algorithm sketched above is, when working at the phenomenological level, at least in principle extendible to arbitrary order in derivatives.  

Despite the fact that hydrodynamics is a foundational component of statistical mechanics, there is still not a generally agreed upon ``derivation" of hydrodynamics from the Schr\"odinger equation.   Hydrodynamics is a dissipative classical theory for a few degrees of freedom, while the many-body Schr\"odinger equation describes linear unitary time evolution on an exponentially large Hilbert space.   Reconciling these two pictures requires a quantitative understanding of how reasonable observers see information loss in the quantum system.   Heuristically, it is well understood that dissipation arises from ``integrating out" the majority of (highly entangled) states in the Hilbert space.    Dephasing of ``off-diagonal" components of the density matrix (in the energy eigenbasis) \cite{deutsch, srednicki} and the exponentially long recurrence times at which this dephasing is ``undone" provide a microscopic picture for thermalization.  Yet this intuition has not yet been turned into a quantitative algorithm for computing the hydrodynamic properties of classical or quantum systems; previous attempts include \cite{mori58, kawasaki, breuer, sasa}.   One could also argue for a quantum kinetic limit \cite{kamenev}, from which hydrodynamics straightforwardly follows.   Yet these prior derivations of hydrodynamics often assume a particular form of Hamiltonian, neglect thermal stochastic effects, or are only valid at weak coupling. 

Given these challenges, recent works \cite{grozdanov, romatschke, rangamani, liu15, jensen} have developed a systematic effective field theory for hydrodynamics.    These recent developments are profound but, in our view, do not  replace a microscopic derivation of hydrodynamics.   Firstly, the action is a theory of spontaneously broken spacetime translation symmetry on the Schwinger-Keldysh contour;  these Goldstone degrees of freedom contain the usual hydrodynamic degrees of freedom, but in a rather abstract way.    Secondly, the effective theories implicitly assume that the only slow degrees of freedom are locally conserved densities and Goldstone bosons, and -- although universally believed to be true --  this assumption has never been formally justified microscopically beyond kinetic theory \cite{kamenev} or gauge-gravity duality \cite{minwalla}.  Finally, these theories are developed in the language of \emph{quantum} field theory, yet it is unclear when (if ever) quantum fluctuations are relevant in hydrodynamics.\footnote{It is clear that quantized `hydrodynamic' fluctuations like phonons can be a crucial part of physics at very low temperatures. Here our focus is on the theory of hydrodynamics as a description of relaxation to thermal equilibrium.   This theory makes sense in highly excited eigenstates, where we will argue that quantum fluctuations are negligible.}    A new microscopic derivation of hydrodynamics may shed light into these questions and provide an interesting perspective on additional subtleties such as fluid frame choices within hydrodynamics, or constraints on higher derivative hydrodynamics beyond the Landau paradigm.

One of the most subtle points in the effective theory for hydrodynamics is the emergence of a thermodynamic entropy which grows over time \cite{liu1612, rangamani18, jensen18}.   In fact, the question of how to define a thermodynamic entropy operator in a many-body quantum system has been debated for many years \cite{robertson, zubarev, ikeda, millis}.   In particular, why does the second law of thermodynamics make sense given the unitarity of quantum mechanics?   Any sensible microscopic derivation of hydrodynamics must  resolve such puzzles, as a key part of hydrodynamic phenomenology is compatibility with the local second law.

Beyond such formalities, a better microscopic understanding of the origin of hydrodynamics will be useful in the development of unbiased numerical algorithms to compute the hydrodynamic coefficients (especially diffusion constants) in microscopic lattice models.   To date, there is no such algorithm that exists: see \cite{altman, wurtz} for heuristic ideas for one dimensional systems.  Because diffusion constants are equivalent to transport coefficients such as electrical conductivity, up to thermodynamic prefactors \cite{kadanoff},  and thermodynamic prefactors are computable numerically \cite{schattner}, an algorithm to compute these diffusion constants is incredibly valuable and could immediately shed light on some of the most challenging problems in condensed matter physics, such as the transport properties of non-Fermi liquids \cite{vandermarel, hussey}.

The purpose of this paper is to start with the many-body Schr\"odinger equation and derive, in an unbiased way, the hydrodynamic theory of energy diffusion, to leading order in the gradient expansion.  More precisely, we will isolate a suitable part of the many-body density matrix: a coarse-grained classical probability distribution for simultaneous measurements of the energy density in every region of space.   We then argue that this classical probability distribution evolves according to a classical Markov process, explicitly identifying the single assumption in the calculation which introduces dissipation -- and from which the entire theory of hydrodynamics follows.   We show that the resulting Markov process is equivalent to nonlinear stochastic hydrodynamics and recover all expected phenomenology, including hydrodynamic cluster decomposition in correlation functions, the fluctuation-dissipation theorem, and the absence of non-hydrodynamic slow degrees of freedom.   Furthermore, we show how the thermodynamic entropy proposed in \cite{millis} quantitatively agrees with the phenomenological hydrodynamic theory of entropy production.  Hence we obtain a more microscopic understanding of the emergence of a local second law of thermodynamics.   We stress that all of these results follow from \emph{one specific postulate} about dissipation in a many-body system.    

The key feature of our framework is the identification of two coarse graining scales, a length $L$ and an energy window $\delta$.  We consider models defined on a very general graph with a well defined dimension, $d$, and short range interactions.  We take $L$ larger than the range of interactions so that the Hamiltonian can be written as
\begin{equation} 
H = \sum_X H(X) + \sum_{X\sim Y} H(X,Y) .  \label{eq:H}
\end{equation} 
$X$ is a set of topologically trivial regions, such that every point on the graph belongs to exactly one $X$.    $X \sim Y$ stands for nearest neighbor regions.\footnote{For simplicity we will take (\ref{eq:H}), which defines a well-posed lattice model, to be exact.   For interactions with range $L_0$, one could imagine corrections to (\ref{eq:H}) of order $\mathrm{e}^{- L/L_0}$, which are still negligible when $L\gg L_0$.} The Hamiltonians $H(X)$ all commute with each other, and, in the absence of other symmetries besides time translation invariance, their joint eigenstates form a complete non-degenerate basis for the Hilbert space.\footnote{We will comment below on other symmetries and on topological defects, which can provide slow degrees of freedom in addition to the energy density.}  Typical eigenvalues of $H(X)$ and $H(X,Y)$ scale as \begin{subequations}\label{eq:HXYscaling}\begin{align}
H(X) &\sim L^d, \\
H(X,Y) &\sim L^{d-1}.
\end{align}\end{subequations}
in units of the fundamental energy scales.  This enables us to treat the interactions between regions as a small perturbation, and forms the basis for our approximations to the full Heisenberg equations of motion.

The splittings between energy density ($ H(X)/ L^d$) eigenvalues range from order 1 to order $\mathrm{e}^{- L^d}$.  We will argue that the dynamics of energy density has a time scale of order $L^{2}$.  Energy differences $\gg L^{-2}$ will dephase rapidly over these hydrodynamic time scales, and can be integrated out in a conventional (Brillouin-Wigner-Wilson) manner.  The tiny energy differences associated with recurrence times are, on the other hand, invisible on the hydrodynamic time scale. We coarse grain energy density into bins of size $\delta$.  This produces a local entropy density, which is the origin of the thermodynamic entropy of the phenomenological theory.  The details of this construction will be provided in the next section.

\section{Setup}
\begin{figure}[t]
\centering
\includegraphics[width=6.5in]{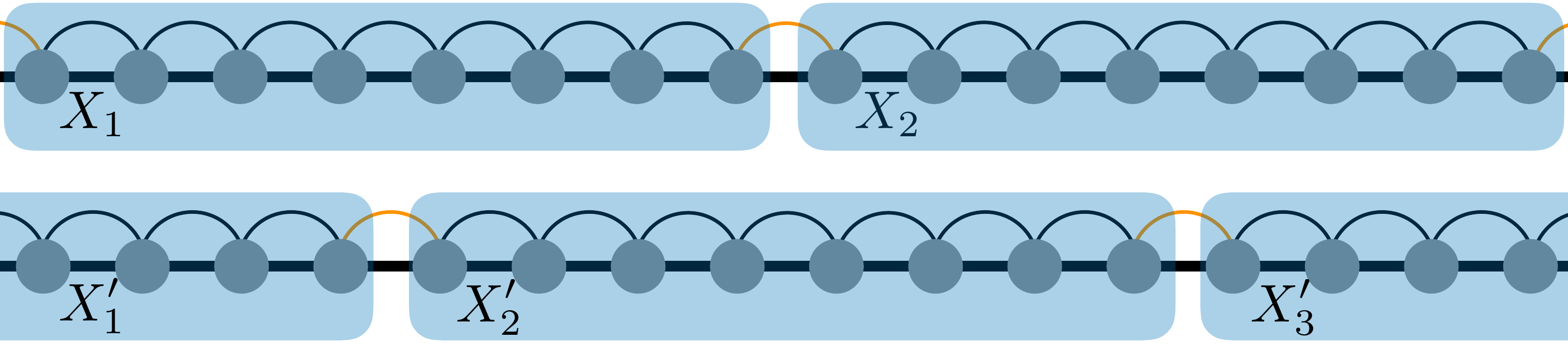}
\caption{A sketch of a many-body quantum system with nearest neighbor interactions on a one dimensional lattice.  We may write $H$ in the form (\ref{eq:H}) in different ways by choosing different disjoint unions of sets $X$ and $X^\prime$, so long as (\ref{eq:Lambda}) is obeyed.  Couplings which are contained in $H(X)$ are black; couplings which are contained in $H(X,Y)$ are orange. }
\label{fig:overlap}
\end{figure}

Consider a many-body quantum system defined on a $d$-dimensional spatial lattice graph $G=(V,\mathcal{E})$;  here $V$ denotes a vertex set, and $\mathcal{E}$ denotes the edge set.    We divide up $V$ into $N\gg 1$ disjoint subsets $X$:  \begin{equation}
V = \bigcup X. \label{eq:Lambda}
\end{equation}We assume that the Hilbert space may be written as a tensor product \begin{equation}
\mathcal{H} = \bigotimes_X \mathcal{H}(X)
\end{equation} and that $\dim(\mathcal{H}(X)) < \infty$.    We assume that the Hamiltonian takes the form (\ref{eq:H}), as sketched in Figure \ref{fig:overlap}.   The regions $X$ each contain of order $L^d$ points of the original vertex set, and $L \gg 1$.  %If we start with finite range couplings on the graph, with range $\ll L$ then couplings between regions that are not adjacent are of order $\mathrm{e}^{-L}$ or smaller and will be negligible in the leading large $L$ expansion that we develop below.

In generic quantum systems, the eigenvalue spectra obey (\ref{eq:HXYscaling}).
The physics that we develop below is insensitive to the precise choice of sets $X$ in (\ref{eq:Lambda}).   We also encourage thinking of $X$ and $Y$ as \emph{points} in a ``superlattice" of regions.   We will often consider a {\it continuum limit} of our equations, which should be valid on length scales much larger than $L$.  In the continuum limit,   $H(X)$ is analogous to the energy density operator (up to a factor of volume).   

We will  consider the following basis for the Hilbert space $\mathcal{H}$.   Let $|\alpha (X)\rangle$ correspond to an eigenvector of $H(X)$ with eigenvalue $E_\alpha (X)$.  (The subscript  $\alpha$ is used to emphasize that this is an individual eigenstate.) Then a complete basis of the Hilbert space is \begin{equation}
\mathcal{H} = \mathrm{span}\left\lbrace \bigotimes_X |\alpha (X)\rangle \right\rbrace \equiv  \mathrm{span} \lbrace | \boldsymbol{\malpha} \rangle\rbrace.
\end{equation}
Here $\boldsymbol{\malpha}$ denotes the collection of all possible eigenstates  $\alpha (X)$.  We will use bold face letters to denote functionals: objects that depend on entire functions $E(X)$ on the superlattice.

We will assume that this basis is non-degenerate, which is equivalent to assuming that there are no more quantities besides energy density which have slow dynamics. Clearly, this is not true if there are other continuous symmetries besides time translation.  We believe that there is a straightforward generalization of our approach that would produce a coupled set of hydrodynamic equations for all conserved currents.  These should be applicable even in phases where gapless Goldstone excitations exist.  Discrete and $p$-form symmetries can introduce other excitations like domain walls and vortices, which will have slow dynamics on the same time scale as the hydrodynamic modes.  We expect that our formalism may be generalized to handle such cases, but will not do so in this paper.  Finally, many-body localized systems \cite{basko, rahul} can have very different long time dynamics and so they require a separate treatment.  The careful reader will note that in one spatial dimension $L^{d-1}$ is not a large number, so some of the claims in the discussion below are not so well justified.  Perhaps this is connected to the fact that many-body localized states and other integrable systems are far more frequent in $d=1$ than $d>1$.

A standard approach to solving Heisenberg's equations of motion is to view them as the evolution of a vector $| A )$ in the real Hilbert space of Hermitian operators $\mathcal{B}$. The {\it Liouvillian} is the operation of commutation with the Hamiltonian:\begin{equation}
\mathcal{L}  = \mathrm{i} [H,\circ].
\end{equation}
We define $\mathcal{L}(X) = \mathrm{i}[H(X),\circ]$ and $\mathcal{L}(X,Y)$ similarly.      A basis for $\mathcal{B}$ is \begin{subequations}\label{eq:orthonormal}\begin{align}
|\boldsymbol{\malpha}\boldsymbol{\malpha} ) &=   |\boldsymbol{\malpha}\rangle\langle \boldsymbol{\malpha}|, \\
|\boldsymbol{\malpha}\boldsymbol{\malpha}^\prime)_+ &=  \frac{ |\boldsymbol{\malpha}\rangle\langle \boldsymbol{\malpha}^\prime| + |\boldsymbol{\malpha}^\prime \rangle\langle\boldsymbol{\malpha}|}{\sqrt{2}},  \\
|\boldsymbol{\malpha}\boldsymbol{\malpha}^\prime)_- &= \frac{ \mathrm{i} |\boldsymbol{\malpha}\rangle\langle \boldsymbol{\malpha}^\prime | -\mathrm{i} |\boldsymbol{\malpha}^\prime\rangle\langle \boldsymbol{\malpha}|}{\sqrt{2}}.
\end{align}\end{subequations}
where  $\boldsymbol{\malpha}\ne \boldsymbol{\malpha}^\prime$:  i.e. there exists an $X$ for which $\alpha(X)\ne \alpha^\prime(X)$.   Using the trace norm as the inner product: \begin{equation}
(A|B) = \mathrm{tr}(A B) 
\end{equation}
along with orthogonality of distinct eigenvectors of each $H(X)$, we see that this is an orthonormal set of Hermitian operators. 

Our goal is now to separate out the eigenvalues of $H(X)$ into bins of size $\delta$.  We take $\delta$ to be a ``small" number obeying $\delta \sim L^{d-\kappa}$ with $0<\kappa< d$, so that in the large $L$ limit $\delta$ is sufficiently small to probe a unique energy density, but $\delta^{-1}$ is also a very short time scale, shorter than the hydrodynamic time scales to which our main equations refer.  As mentioned in the introduction, there are extremely tiny energy differences in the spectrum of $H(X)$, which are impossible to resolve over hydrodynamic time scales.  States within a $\delta$ band are essentially degenerate and define an energy dependent local entropy within each block.

Now define a projection operator $\mathbb{P}(\mathbf{E})$ onto a given band of eigenvalues in each $X$, which we will also write as $|\mathbf{E})$:
\begin{equation}
|\mathbf{E}) = \sum_{\boldsymbol{\malpha}}  \mathrm{\Theta}(E_\alpha(X) - E(X))\mathrm{\Theta}(E(X)+ \delta - E_\alpha(X) ) |\boldsymbol{\malpha}\boldsymbol{\malpha} ) . 
\end{equation}
The operators $|\mathbf{E})$ are defined on a hypercubic lattice in $N$-dimensional energy space -- the spacing between each lattice point is $\delta$.   All $\mathbb{P}(\mathbf{E})$ are diagonal in the same $\boldsymbol{\malpha}$ basis.
 We also define \begin{equation}
\Omega(\mathbf{E}) =  \sum_{\boldsymbol{\malpha}} \prod_{X} \frac{  \mathrm{\Theta}(E_\alpha(X) - E(X))\mathrm{\Theta}(E(X)+ \delta - E_\alpha(X) )}{\delta} = \prod_X \omega(E(X),X) = \prod_X \mathrm{e}^{S(E(X),X)},  \label{eq:Omega}
\end{equation}
which is related to the norm of $|\mathbf{E})$: 
\begin{equation}
(\mathbf{E}|\mathbf{E}^\prime) = \mdelta(\mathbf{E}-\mathbf{E}^\prime) \times \delta^N \Omega(\mathbf{E}),
\end{equation}
Finally, we define the projector the subspace of operators that are diagonal in the energy basis: \begin{equation}
\mathcal{D} = \sum_{\mathbf{E}}  \frac{|\mathbf{E})(\mathbf{E}|}{\delta^N\Omega(\mathbf{E})} .
\end{equation}
With these notational conventions set, we can proceed to analyze the large $L$ limit of the equations for such diagonal operators.  We will find that, to leading order in $L$, the equations close.

\section{Emergence of Dissipation}\label{sec:diss}
The question that we will address in this paper is the time evolution of $\mathcal{D}\rho(t)$ for generic interacting quantum systems in highly excited states -- here $\rho(t)$ is the time-evolved density matrix.    It is instructive to define \begin{equation}
(\mathbf{E}|\rho(t)) = \delta^N p(\mathbf{E},t),
\end{equation}
such that \begin{equation}
\mathcal{D}|\rho(t)) = \delta^N \sum_{\mathbf{E}} p(\mathbf{E},t) |\mathbf{E}).  \label{eq:DrhoE}
\end{equation}
We can think of $p(\mathbf{E})$ as a many-body probability distribution function for simultaneous measurements of the local energy density operators $H(X)$.    It is a \emph{classical} approximation to the quantum mechanical state of the system.  In the continuum limit, this distribution should be thought of as a functional of the energy density.   $p(\mathbf{E})$ contains exponentially many degrees of freedom, but this number is also exponentially smaller than the total number of eigenstates in the system.    Dissipation will arise from integrating out nearly all of the information contained within the density matrix and the emergence of simple Markovian dynamics for $p(\mathbf{E})$.  The main purpose of this section is to derive this stochastic process, elucidating the necessary assumptions as we go. 

\subsection{Memory Matrix Formalism}
Our starting point is the microscopic many-body Schr\"odinger equation \begin{equation}
\frac{\mathrm{d}}{\mathrm{d}t} |\rho(t)) = -\mathcal{L}|\rho(t)).   
\end{equation}
So far, the time evolution is unitary (i.e. $\mathcal{L}$ is a real antisymmetric matrix) -- there is no dissipation.   However, our goal is not to compute the full $|\rho(t))$, as such a task is generically not possible.   Instead, our goal is to only compute the time evolution of $p(\mathbf{E},t)$ -- the classical probability of finding the system at a given energy (in every region $X$).

This can be accomplished using the memory function formalism \cite{zwanzig, mori, forster, lucasbook}.   The basic objective is to start from the following matrix identity, with $\mathcal{N} = 1-\mathcal{D}$:  \begin{equation}
\mathcal{D}\mathrm{e}^{-\mathcal{L}t}\mathcal{D} = \mathcal{D} - \int\limits_0^t \mathrm{d}s \mathcal{D}\mathrm{e}^{-\mathcal{LN}(t-s)}\mathcal{LD}\mathrm{e}^{-\mathcal{L}s}\mathcal{D}. \label{eq:DeLtD}
\end{equation} This is the integrated form of the Schwinger-Karplus identity for the derivative of an operator function.  

We have used the fact that $\mathcal{ND}=0$ to simplify terms above, together with the identity \begin{equation}
\mathcal{DLD} = \sum_{\mathbf{E},\mathbf{E}^\prime}  (\mathbf{E}|\mathcal{L}|\mathbf{E}^\prime) \frac{ |\mathbf{E}) (\mathbf{E}^\prime|}{\delta^{2N}\Omega(\mathbf{E})\Omega(\mathbf{E}^\prime)}= 0,
\end{equation}
which can be derived using the fact that \begin{equation}
(\mathbf{E}|\mathcal{L}|\mathbf{E}^\prime) = \mathrm{tr}(\mathbb{P}(\mathbf{E})[H,\mathbb{P}(\mathbf{E}^\prime)]) = \mathrm{tr}(H[\mathbb{P}(\mathbf{E}^\prime),\mathbb{P}(\mathbf{E})]) = 0.  \label{eq:trcycle}
\end{equation}
Let us choose an initial condition such that $\mathcal{N}|\rho(0))=0$.
From (\ref{eq:DeLtD}) we obtain : \begin{equation}
\frac{\mathrm{d}}{\mathrm{d}t} \mathcal{D}|\rho(t)) = \int\limits_0^t \mathrm{d}s \; \mathcal{DLN} \mathrm{e}^{-\mathrm{i}\mathcal{NLN}s} \mathcal{NLD}|\rho(s))  \equiv -\int\limits_0^t \mathrm{d}s \; \mathcal{M}(t-s) \mathcal{D}|\rho(s)). \label{eq:MM}
\end{equation}
We have thus completely integrated out the degrees of freedom $\mathcal{N}|\rho(t))$, which will become non-trivial for $t>0$.   The price for this is the non-locality of (\ref{eq:MM}).

However, we are naturally interested in the limit where each region $X$ contains a reasonable number of degrees of freedom.   This leads to a number of important simplifications.   Firstly, if each region $X$ consists of $L$ sites in each direction (and so contains $\sim L^d$ sites in  total), then $H_I$ contains $\exp[L^d]$ eigenvalues which will generically be exponentially close together (so long as we avoid the minimal or maximal eigenvalue).     Secondly, by construction \begin{equation}
\mathcal{L}(X) |\mathbf{E}) = \sum_{|\boldsymbol{\malpha}\rangle\notin \mathrm{ker}(\mathbb{P}(\mathbf{E}))}  (E_\alpha (X)-E_\alpha(X))|\boldsymbol{\malpha\malpha}) = 0.
\end{equation}
Therefore $\mathcal{DLN}$ and $\mathcal{NLD}$ receive contributions only from $\mathcal{L}(X,Y)$.   In contrast, \begin{equation}
\mathcal{L}(X)|\boldsymbol{\malpha\malpha}^\prime)_\pm = \pm (E_\alpha (X)-E_\alpha^\prime(X)) |\boldsymbol{\malpha\malpha}^\prime)_\mp + \cdots,  \label{eq:NLN}
\end{equation}
where $\cdots$ denotes contributions which are subleading in $L$;  thus $\mathcal{NLN}$ can receive contributions from $\mathcal{L}(X)$.  Finally, (\ref{eq:HXYscaling}) suggests that $\mathcal{L}(X,Y)$ contributes a small factor of $L^{-1}$ and can be ignored in $\mathcal{NLN}$ but not in $\mathcal{DLN}$ or $\mathcal{NLD}$.   In other words, (\ref{eq:MM}) describes the slow dynamics of $|\mathbf{E})$ transitioning among themselves.   The dynamics is slow both because of the overall prefactor of $L^{-2}$, and because the transitions between $|\mathbf{E})$ arise through ``fast modes" $|\boldsymbol{\malpha\malpha}^\prime)_\pm$ which we have integrated out and which (relative to one another) dephase extremely quickly.

To quantify how dephasing leads to dissipation, we explicitly evaluate $(\mathbf{E} |\mathcal{M}(t)|\mathbf{E}^\prime)$ to leading order in $L$.   We start from
 $\mathcal{NLD}|\mathbf{E})$:
\begin{align}
\mathcal{NLD}|\mathbf{E}) &= \sum_{X\sim Y} |\mathrm{i}[H(X,Y), \mathbb{P}(\mathbf{E})]) \notag \\
&= \sqrt{2} \sum_{X\sim Y}  \sum_{\boldsymbol{\mgamma}} \sum_{|\boldsymbol{\malpha}\rangle \notin \mathrm{ker}(\mathbb{P}(\mathbf{E}))} \mathrm{Re}(\langle \boldsymbol{\mgamma}|H(X,Y)|\boldsymbol{\malpha}\rangle) |\boldsymbol{\malpha\mgamma})_- +\mathrm{Im}(\langle \boldsymbol{\mgamma}|H(X,Y)|\boldsymbol{\malpha}\rangle) |\boldsymbol{\malpha\mgamma})_+ .\label{eq:alphabeta}
\end{align}
There are a number of important comments to make about (\ref{eq:alphabeta}).   (\emph{i}) 
Using the same manipulation as (\ref{eq:trcycle}) we can confirm that $(\boldsymbol{\malpha\malpha}|\mathcal{L}|\mathbf{E}) = 0$ for any $\mathbf{E}$ and $\boldsymbol{\malpha}$.  (\emph{ii}) In (\ref{eq:alphabeta}), $\alpha(Z)=\gamma(Z)$ if and only if $Z\ne X,Y$ since $H(X,Y)$ does not act on $\mathcal{H}_Z$.      (\emph{iii}) We may restrict the sum on $\boldsymbol{\mgamma}$ in (\ref{eq:alphabeta}) to $\mathbb{P}(\mathbf{E})|\boldsymbol{\mgamma}\rangle = 0$, since $[\mathbb{P}(\mathbf{E}), |\boldsymbol{\malpha}\rangle\langle\boldsymbol{\mgamma}|]=0$ and so such terms may be cancelled in the first line of (\ref{eq:alphabeta}).   Next, we observe that $(\mathbf{E}|\mathcal{DLN} = -(\mathcal{NLD}|\mathbf{E}))^\mathsf{T}$ since $\mathcal{L}$ is antisymmetric.   At leading order in $L$ (and $\delta$), we saw in (\ref{eq:NLN}) that $\mathcal{NLN}$ is diagonal.   Hence we conclude that $(\mathbf{E}|\mathcal{M}(t)|\mathbf{E}^\prime) \ne 0$ only when $\mathcal{NLD}|\mathbf{E})$ and $\mathcal{NLD}|\mathbf{E}^\prime)$ contain identical operators $|\boldsymbol{\malpha\mgamma})_\pm$.    From point (\emph{iii}) above, we conclude that there are only two possible contributions to $(\mathbf{E} |\mathcal{M}(t)|\mathbf{E}^\prime)$.   Firstly, if $\mathbf{E}=\mathbf{E}^\prime$, then we sum over all possible transitions from any $|\boldsymbol{\malpha}\rangle \notin\mathrm{ker}(\mathbb{P}(\mathbf{E}))$ to any $|\boldsymbol{\mgamma}\rangle \in\mathrm{ker}(\mathbb{P}(\mathbf{E}))$;   if $\mathbf{E}\ne \mathbf{E}^\prime$, we sum over all possible transitions from any $|\boldsymbol{\malpha}\rangle \notin\mathrm{ker}(\mathbb{P}(\mathbf{E}))$ to any $|\boldsymbol{\mgamma}\rangle \notin\mathrm{ker}(\mathbb{P}(\mathbf{E}^\prime))$.    This is neatly summarized in the final result \begin{subequations}\label{eq:calM}\begin{align}
(\mathbf{E}|\mathcal{M}(t)|\mathbf{E}^\prime) &= -2\sum_{X\sim Y}\sum_{\substack{|\boldsymbol{\malpha}\rangle \notin \mathrm{ker}(\mathbb{P}(\mathbf{E})) \\ |\boldsymbol{\mgamma}\rangle \notin \mathrm{ker}(\mathbb{P}(\mathbf{E}^\prime))}} |\langle \boldsymbol{\mgamma}|H(X,Y)|\boldsymbol{\malpha}\rangle|^2 \mathrm{e}^{-\mathrm{i}(E_\alpha(X)+E_\alpha(Y)-E_\gamma(X)-E_\gamma(Y))t}, \;\;\;\;\; (\mathbf{E}\ne\mathbf{E}^\prime), \\
(\mathbf{E}|\mathcal{M}(t)|\mathbf{E}) &= 2\sum_{\mathbf{E}^\prime \ne \mathbf{E}} \sum_{X\sim Y}\sum_{\substack{|\boldsymbol{\malpha}\rangle \notin \mathrm{ker}(\mathbb{P}(\mathbf{E})) \\ |\boldsymbol{\mgamma}\rangle \in \mathrm{ker}(\mathbb{P}(\mathbf{E}^\prime))}} |\langle \boldsymbol{\mgamma}|H(X,Y)|\boldsymbol{\malpha}\rangle|^2 \mathrm{e}^{-\mathrm{i}(E_\alpha(X)+E_\alpha(Y)-E_\gamma(X)-E_\gamma(Y))t}.
\end{align}\end{subequations}
The phase factor in the exponential comes from a simple application of (\ref{eq:NLN}).

%  These modes have a lifetime $\sim L^2$ and do not exist in the absence of $H(X,Y)$.  We will return to the consequences of such hydrodynamic modes in Section \ref{sec:higherderivative}, at least in a highly simplified context.    Our conclusion is that (\ref{eq:calM}) is a quantitatively correct description of physics on length scales sufficiently long compared to $L$.  In a generic setting, we expect that (\ref{eq:calM}) is a qualitatively (but not quantitatively) accurate description of physics even on length scales comparable to $L$ (which serves as a hard cutoff in our approach).  Unless stated otherwise, we will henceforth assume that (\ref{eq:calM}) holds.

Recall from (\ref{eq:MM}) that our final formula for the rate of change of $\mathcal{D}|\rho(t))$ depends on an integral over $\mathcal{M}$ for all times.  Given our choice of $\delta^{-1}$ vanishing with $L$, we approximate that on any finite, $L$-independent time scale:  
\begin{equation}
\int\limits_0^t \mathrm{d}s\; \mathrm{e}^{-\mathrm{i}(E_\alpha(X)+E_\alpha(Y)-E_\gamma(X)-E_\gamma(Y))t} \approx \mpi \; \mdelta(E_\alpha(X)+E_\alpha(Y)-E_\gamma(X)-E_\gamma(Y)).  \label{eq:ergodic}
\end{equation}
Rapid dephasing will cancel all contributions in (\ref{eq:calM}) unless we have \emph{local energy conservation}.   Combining (\ref{eq:MM}), (\ref{eq:calM}) and (\ref{eq:ergodic}) we obtain a \emph{classical Markov process} which describes the spatial dynamics of energy \emph{including all statistical fluctuations}:  \begin{equation}
\delta^N \partial_t p(\mathbf{E},t) = \delta^{2N} \sum_{\mathbf{E}^\prime \ne \mathbf{E}} W(\mathbf{E},\mathbf{E}^\prime) \left(\frac{p(\mathbf{E}^\prime,t)}{\Omega(\mathbf{E}^\prime)} - \frac{p(\mathbf{E},t)}{\Omega(\mathbf{E})}\right)
\end{equation}
where \begin{equation}
W(\mathbf{E},\mathbf{E}^\prime) = \frac{1}{\delta^{2N}} \times 2\mpi \sum_{X\sim Y}\sum_{\substack{|\boldsymbol{\malpha}\rangle \notin \mathrm{ker}(\mathbb{P}(\mathbf{E})) \\ |\boldsymbol{\mgamma}\rangle \notin \mathrm{ker}(\mathbb{P}(\mathbf{E}^\prime))}}|\langle \boldsymbol{\mgamma}|H(X,Y)|\boldsymbol{\malpha}\rangle|^2\mdelta(E_\alpha(X)+E_\alpha(Y)-E_\gamma(X)-E_\gamma(Y)) . \label{eq:W}
\end{equation}
 Note that $W(\mathbf{E},\mathbf{E}^\prime) = W(\mathbf{E}^\prime,\mathbf{E})$, and that the factor of $\delta$ is chosen so that $W$ is invariant under rescalings of $\delta$ (when $\delta$ is sufficiently small and the spectrum of $H(X)$ is continuous).    It is now simple to take the continuum limit in energy space:  \begin{equation}
 \partial_t p(\mathbf{E},t) = \int \mathrm{d}\mathbf{E}^\prime \; W(\mathbf{E},\mathbf{E}^\prime) \left(\frac{p(\mathbf{E}^\prime,t)}{\Omega(\mathbf{E}^\prime)} - \frac{p(\mathbf{E},t)}{\Omega(\mathbf{E})}\right)
 \label{eq:markov}
 \end{equation}
 This continuous time Markov process on a functional space is our first key result.

There is an important subtlety in the above arguments.  There are slow, hydrodynamic modes with wavelengths $\sim \frac{1}{2}L$ which we have implicitly ignored  when we assumed $\mathcal{NLN}$ was independent of $\mathcal{L}(X,Y)$.   These modes lead to corrections to the Markov process above.  We will address this issue in a more quantitative way in Section \ref{sec:higherderivative}, and show that such corrections matter only on the shortest time scales accessible in (\ref{eq:markov}).   
 
As our primary interest in (\ref{eq:markov}) is on the late time limit of the dynamics, we will assume the Markov equation from here on.  There are two immediate physical observations that we can make.  Firstly, the transition rates $W(\mathbf{E},\mathbf{E}^\prime)$ are those predicted by Fermi's golden rule.   So (\ref{eq:markov}) admits a very simple physical interpretation:  a conserved energy ($\sum E(X)$) is exchanged between adjacent sites.   The energy coming from the boundary degrees of  freedom is smaller by a power of $L$, and so can be neglected to leading order in $L$.   Furthermore, as we will prove explicitly later, the only equilibria are of the form \begin{equation}
p_{\mathrm{eq}}(\mathbf{E}) = \Omega(\mathbf{E}) \mathcal{F}\left(\sum_X E(X)\right).  
\end{equation}
The dynamics relaxes to the microcanonical ensemble;  the function $\mathcal{F}$ is determined by the initial conditions in a straightforward way: \begin{equation}
\mathcal{F}(E_{\mathrm{tot}}) =\dfrac{\displaystyle  \int\mathrm{d}\mathbf{E}\; \mdelta\left(E_{\mathrm{tot}} - \sum_X E(X)\right)p(\mathbf{E},0)}{\displaystyle  \int\mathrm{d}\mathbf{E}\; \mdelta\left(E_{\mathrm{tot}} - \sum_X E(X)\right)\Omega(\mathbf{E})}  \label{eq:calF}
\end{equation}
 
 We expect that (\ref{eq:ergodic}) -- and thus (\ref{eq:markov}) --  is a sensible approximation for generic many-body systems in a highly excited state with finite (neither maximal or minimal) energy density.   However, we reiterate that there are a number of important cases where this approximation will fail, which we list for completeness.   One possibility is that there are additional local conservation laws, such as charge or spin.  It is natural to generalize the above formalism to account for such conserved charges (so long as the number is finite) and we will not do so explicitly in this paper.  In integrable models the list of conservation laws should be much larger, and it would be interesting to compare this formalism to ``generalized hydrodynamics" \cite{doyon, joelmoore, denardis}.    Topological defects such as vortices in a superfluid can only decay by colliding with other defects, and so the defect motion becomes an additional slow degree of freedom \cite{halperin}.\footnote{Actually, vortices are related to a topological two form current \cite{seiberg} (an ordinary current for $d = 2$) and so are taken into account by a hydrodynamics \cite{iqbal} including all the conserved quantities of the theory.}  There are also defects associated with spontaneoulsy broken discrete symmetries whose motion we have not accounted for.   Finally, in a many-body localized (MBL) phase, quantum ergodicity is simply broken;  the emergent spectrum of $H(X)$ should appear sufficiently discrete so as to destroy the Markovian limit \cite{basko, rahul}.    We expect that our formalism could be generalized to study symmetry broken phases, but not many-body localization.  One way of understanding MBL is that it occurs in systems that have an infinite number of effectively local conservation laws \cite{serbyn}.   Such conservation laws are local in position, not momentum, space, in contrast to integrable models such as free theories.  We leave a more quantitative understanding of defect motion and MBL phases for future work.   It would be interesting if (\ref{eq:ergodic}) is equivalent to the eigenstate thermalization hypothesis \cite{deutsch, srednicki}: we are not aware of any theory where one should hold without the other.

\subsection{The Functional Fokker-Planck Equation}
So far we have used the large $L$ limit to simplify the dynamics so as to see an emergent energy conservation.   However, the large $L$ limit \emph{also} guarantees that the resulting dynamics is slow, and our next goal is to exploit the slow dynamics in $\mathbf{E}$-space to simplify (\ref{eq:markov}) to a high-dimensional diffusion equation in $\mathbf{E}$-space:  a  functional Fokker-Planck equation (FFPE).    

The key observation is that for a local Hamiltonian, $H(X,Y)|\boldsymbol{\malpha}\rangle$ leads to a superposition of states $|\boldsymbol{\mgamma}\rangle$ where $E_\alpha(X)-E_\gamma(Y) \sim E_\gamma(X)-E_\gamma(Y)\sim L^0$.   To see this, consider the following manipulations:  \begin{equation}
\left\langle \boldsymbol{\malpha} \left| \left(\mathcal{L}_X^n  H(X,Y) \right)\right| \boldsymbol{\mgamma}\right\rangle =  (E_\alpha(X)-E_\gamma(X))^n \langle \boldsymbol{\malpha}|H(X,Y)|\boldsymbol{\mgamma}\rangle = \langle \boldsymbol{\malpha}|\mathcal{O}^{(n)}_Y(X)|\boldsymbol{\mgamma}\rangle \propto n L^{d-1}.   \label{eq:L0bound}
\end{equation}
here $\mathcal{L}_X H(X,Y) = [H(X),H(X,Y)]$ and $\mathcal{O}^{(n)}_Y(X)$ denotes an operator which acts in $\mathcal{H}_X$ but has support within $n$ sites of the boundary with $Y$.   Taking $n\sim L$, (\ref{eq:L0bound}) implies  that in the large $L$ limit, $E_\alpha(X)-E_\gamma(X) \propto L^0$.    A heuristic understanding of this fact follows from the observation that $H(X,Y)$ is a sum of local operators.   Each local operator can only shift the energy by a constant because it only affects the wave function on a small number of sites.   (\ref{eq:L0bound}) makes this precise.

Now let us revisit (\ref{eq:markov}), which describes a Markov process with transitions between states of different energy.  From (\ref{eq:W}), we see that these transitions occur between states with very similar energy.   Therefore it is natural to expand (\ref{eq:markov}) in the small parameter $\boldsymbol{\mepsilon} =  \mathbf{E}^\prime - \mathbf{E}$.    Firstly, observe that
\begin{equation}
\frac{p(\mathbf{E}^\prime,t)}{\Omega(\mathbf{E}^\prime)} - \frac{p(\mathbf{E},t)}{\Omega(\mathbf{E})} = \epsilon(X)\partial_{E(X)}\left(\frac{p(\mathbf{E})}{\Omega(\mathbf{E})}\right) + \frac{1}{2}\epsilon(X)\epsilon(Y)\partial_{E(X)}\partial_{E(Y)}\left(\frac{p(\mathbf{E})}{\Omega(\mathbf{E})}\right) + \cdots,  \label{eq:pepsilonexpansion}
\end{equation}
Secondly, it is useful to use the exact identity $W(\mathbf{E},\mathbf{E}^\prime) = W(\mathbf{E}^\prime,\mathbf{E})$ to expand \begin{equation}
W(\mathbf{E},\mathbf{E}  + \boldsymbol{\mepsilon}) = \left[W^{(0)}\left(\mathbf{E}+ \frac{\boldsymbol{\mepsilon}}{2}\right) + \sum_{X,Y} W^{(2)}\left(\mathbf{E}+ \frac{\boldsymbol{\mepsilon}}{2},X,Y\right)\epsilon(X)\epsilon(Y) +\cdots\right] \times g(\boldsymbol{\mepsilon}),
\end{equation} 
where $g$ is a sharply peaked function near $\boldsymbol{\mepsilon} = \mathbf{0}$ which is even in all $\epsilon(X)$.     Integrating over $\mathrm{d}\mathbf{E}^\prime = \mathrm{d}\boldsymbol{\mepsilon}$ in (\ref{eq:markov}), and using (\ref{eq:pepsilonexpansion}), the leading order contribution will be $\mathrm{O}(\epsilon^2)$ as $\boldsymbol{\mepsilon} \rightarrow \mathbf{0}$.  Hence we find that (\ref{eq:markov}) can be effectively approximated by a Functional Fokker Planck Equation (FFPE):
\begin{equation} 
\partial_t p(\mathbf{E},t) = \sum_{X,Y} \partial_{E(X)} \left(\Sigma(\mathbf{E},X,Y)\partial_{E(Y)} \left(\frac{p(\mathbf{E},t)}{\Omega(\mathbf{E})}\right)\right) + \cdots,  \label{eq:FPE}
\end{equation}
where \begin{equation}
\Sigma(\mathbf{E},X,Y) = \frac{1}{2} \int \mathrm{d}\boldsymbol{\mepsilon} \; \epsilon(X)\epsilon(Y) W(\mathbf{E},\mathbf{E}+\boldsymbol{\mepsilon}) = \frac{1}{2} \int \mathrm{d}\boldsymbol{\mepsilon} \; \epsilon(X)\epsilon(Y) g(\boldsymbol{\mepsilon})W^{(0)} \left(\mathbf{E}+ \frac{\boldsymbol{\mepsilon}}{2}\right) 
\end{equation}
and the $\cdots$ in (\ref{eq:FPE}) denotes terms that are subleading in $L$:  since $\epsilon \sim L^0$, only the lowest order term in $\mathbf{E}$-derivatives contributes at leading order in $L$.    For later convenience, we will also define \begin{equation}
\mathcal{D}(\mathbf{E},X,Y) = \frac{\Sigma(\mathbf{E},X,Y)}{\Omega(\mathbf{E})}.  \label{eq:calD}
\end{equation}
Note that from (\ref{eq:W}),  $\mathcal{D}(\mathbf{E},X,Y)$ depends on only $E(X)$ and $E(Y)$.    

Let us now discuss the scaling of $\mathcal{D}$ with $L$.   We start with (\ref{eq:W}).  Note that there exist (generic) transitions for which
\begin{equation}
\langle \boldsymbol{\mgamma}|H(X,Y)|\boldsymbol{\malpha}\rangle \sim L^{d-1};
\end{equation}
despite the fact that the energy exchange $\epsilon \sim L^0$ between $|\boldsymbol{\mgamma}\rangle$ and $|\boldsymbol{\malpha}\rangle$ is small, there are $L^{d-1}$ possible terms that can exchange energy.    The sum over states in (\ref{eq:W}) leads to an overall factor of $\delta^{2N}\Omega(\mathbf{E})$;  these factors are cancelled by the $\delta^{-2N}$ prefactor in (\ref{eq:W}), and the $\Omega^{-1}$ prefactor in (\ref{eq:calD}).   Finally, the $\mdelta$ function in (\ref{eq:W}) leads to a factor of $L^{-d}$. This scaling is easiest to understand by interpreting (\ref{eq:W}) as an integral.    Putting all of this together, we conclude that \begin{equation}
\mathcal{D}\sim L^{d-2}.  \label{eq:DLd2}
\end{equation}
As we will see, the prefactor of $-2$ above is intimately related to the emergence of diffusive hydrodynamics.

(\ref{eq:FPE}) is our final and main result of this section.   It demonstrates that the energy dynamics in an arbitrary ergodic many-body quantum system is effectively described by a continuous diffusion equation on a high dimensional space.   Taking the continuum limit and thinking of $\mathbf{E}$ as parameterizing smooth functions $E(X)$, we obtain a functional diffusion equation.\footnote{Functional diffusion equations have been studied in a very different context in the exact renormalization group \cite{polchinski}.}

For simplicity, in the sections that follow, we will assume discrete translation invariance :  $\omega$ (defined in (\ref{eq:Omega})) is \begin{equation}
\omega(E(X),X) =  \omega(E(X)).
\end{equation}
Using the textbook definition of inverse temperature $\beta$ in the microcanonical ensemble: \begin{equation}
\partial_{E} \log \omega(E) = \beta(E).   \label{eq:betaE}
\end{equation}
We also assume that \begin{equation}
\mathcal{D}(\mathbf{E},X,Y) =  \mathcal{D}(E(X),E(Y)) \Lambda(X,Y)   \label{eq:DLambda}
\end{equation}
where $\Lambda(X,Y)$ is the discretized Laplacian $-\nabla^2$ on the ``superlattice" graph of regions $X$:  \begin{equation}
\Lambda(X,Y) = \left\lbrace \begin{array}{ll} \displaystyle \sum_{Z\sim X} 1 &\   Y= X \\ -1 &\  X\sim Y \end{array}\right..  \label{eq:Laplacian}
\end{equation}
 Observe that $\Lambda(X,Y)$ is, as a matrix, independent of the size $L$ of the unit cell of the superlattice.  In terms of physical coordinates, we conclude that the continuous space limit of $\Lambda(X,Y)$ is $-L^2 \nabla^2 \mdelta(X-Y)$.   We take $\mathcal{D}(E(X),E(Y)) = \mathcal{D}(E(Y),E(X))$.  
The resulting FFPE is invariant under discrete translations and all inhomogeneity in the system is encoded in the initial conditions $p(\mathbf{E},0)$.   We emphasize that our formalism can deal equally well with inhomogeneous systems, but it is difficult to analyze all of the different forms that inhomogeneity could take in a single discussion.

\section{Hydrodynamics}\label{sec:hydro}
One way to interpret (\ref{eq:FPE}) is simply to write the FFPE as a nonlinear stochastic equation in the Stratonovich calculus \cite{stochastic}.    In this framework, one finds that the stochastic differential equation \begin{equation}
\partial_t A_a(t) = F_a(A(t)) + \sigma_{a\alpha}(A) \xi_\alpha(t),
\end{equation}
where $\xi_\alpha(t)$ is uncorrelated Gaussian white noise: \begin{equation}
\overline{\xi_\alpha(t)\xi_\beta(s)} = \mdelta_{\alpha\beta} \mdelta(t-s),
\end{equation}
is equivalent to the Fokker-Planck equation \begin{equation}
\partial_t p(A, t) = \sum_a \frac{\partial}{\partial A_a} \left(-F_a(A) p(A,t) + \frac{1}{2}\sum_{b,\alpha} \sigma_{\alpha a}(A)\frac{\partial}{\partial A_b} \left(\sigma_{b\alpha}(A) p(A,t)\right)\right)  \label{eq:stratonovichFPE}
\end{equation}
Hence, we need to identify a sufficient number of independent Gaussian noise variables such that the FFPE (\ref{eq:FPE}) can be recast as a stochastic partial differential equation.  

In order to accomplish this, we first rewrite (\ref{eq:FPE}) in a slightly different form:  \begin{equation}
\partial_t p =- \sum_X \partial_{E(X)} \left(\left(\sum_Y \mathcal{D}(\mathbf{E},X,Y) \partial_{E(Y)} \log \omega(E(Y))\right)p - \sum_Y \mathcal{D}(\mathbf{E},X,Y) \partial_{E(Y)} p  \right)
\end{equation}
 A natural choice is to put a noise variable on every edge $\mathfrak{e}=(X,Y)$ (with $X\sim Y$) in the superlattice, and choose noise strength \begin{equation}
\sigma(\mathbf{E}, Z, \mathfrak{e}) = \left\lbrace \begin{array}{ll} 0 &\ Z\ne X,Y \\ \sqrt{2\mathcal{D}(\mathbf{E},X,Y)} &\ Z = X \\  -\sqrt{2\mathcal{D}(\mathbf{E},X,Y)} &\ Z = X \end{array}\right.. 
\end{equation}
From (\ref{eq:DLambda}) and (\ref{eq:stratonovichFPE}), we conclude that \begin{equation}
\sum_{\mathfrak{e}} \sigma(\mathbf{E}, X, \mathfrak{e})  \sigma(\mathbf{E}, Y, \mathfrak{e})  =  2 \mathcal{D}(\mathbf{E},X,Y),
\end{equation}
Consequently these noise variables will reproduce the second derivative term in (\ref{eq:FPE}).   To reproduce the first derivative term, observe that (if $\langle \cdots \rangle$ denotes averaging over noise): \begin{equation}
\partial_t \langle E(X)\rangle = \sum_Y \left\langle \mathcal{D}(\mathbf{E},X,Y) \beta(E(Y)) - (\partial_{E(X)} - \partial_{E(Y)}) \mathcal{D}(\mathbf{E},X,Y) \right\rangle 
\end{equation} 
Due to the $\mathbf{E}$-derivatives, the latter term above is suppressed by a factor of $L^{-d}$.

We conclude that to leading order in  $L$, (\ref{eq:FPE}) is equivalent to \begin{equation}
\partial_t E(X) =  \sum_{X,Y} \mathcal{D}(\mathbf{E}, X,Y) \beta(E(Y,t))+ \xi(X,t),  \label{eq:stochastichydro}
\end{equation}
where $\xi(X,t)$ is Gaussian white noise, with zero-mean, and variance
\begin{equation}
\overline{\xi(X,t)\xi(X^\prime,t^\prime)} = 2\mathcal{D}(\mathbf{E},X,X^\prime)\mdelta(t-t^\prime) .  \label{eq:noiseD}
\end{equation}
This stochastic equation is interpreted in the Stratonovich calculus.   (\ref{eq:stochastichydro}) is stochastic nonlinear hydrodynamics, describing the diffusion of energy across the lattice.  Our main result is the unambiguous derivation of (\ref{eq:stochastichydro}) from the microscopic Schr\"odinger equation together with the specific Markov approximation (\ref{eq:markov}) described above.   As we now show, this Markov approximation is sufficient to recover the complete statistical mechanical theory of hydrodynamics with classical noise. 

Let us first begin by neglecting the noise $\xi(X,t)$ in (\ref{eq:stochastichydro}).   
Taking the continuum limit of the discrete Laplacian $\Lambda(X,Y)$, we obtain \begin{equation}
\partial_t E = -\nabla \cdot \left(\mathcal{D}(E) L^2 \nabla \beta(E)\right),  \label{eq:diffusioneq}
\end{equation}
which is the nonlinear energy diffusion equation predicted by hydrodynamics.   As we will see in Section \ref{sec:higherderivative}, the approach developed in this paper is not sophisticated enough to quantitatively capture higher derivative corrections to (\ref{eq:diffusioneq}), which is why we have taken $\mathcal{D}(E) = \mathcal{D}(E(X)=E, E(Y)=E)$ to depend only a single energy variable.   

Near equilibrium, we may approximate $\mathcal{D} \approx \mathcal{D}(\mathbf{E})$.  Let $\bar E(\beta)$ be the inverse of the function $\beta(E)$, and recall that the specific heat $C = -\beta^2 \partial_\beta \bar E$.   Using \begin{equation}
\nabla \beta(E) = -\frac{\beta^2}{C} \nabla E,
\end{equation}
we conclude that near equilibrium the average energy $E$ obeys a diffusion equation with diffusion constant \begin{equation}
D_{\mathrm{e}} = \frac{\mathcal{D}L^2 \beta^2}{C}.  \label{eq:Denergy}
\end{equation}
From (\ref{eq:DLd2}), along with thermodynamic extensivity ($C\sim L^d$), we observe that $D_{\mathrm{e}}$ is independent of the artificial cutoff $L$.   $D_{\mathrm{e}}$ is the physical diffusion constant of energy at a given temperature, and can be measured by two-point thermal correlators \cite{kadanoff}. 

Next let us address the presence of noise.   Using (\ref{eq:betaE}), (\ref{eq:noiseD}), and (\ref{eq:Denergy}), we  conclude that near equilibrium, the noise strength obeys the fluctuation-dissipation theorem:
\begin{equation}
2\mathcal{D}(X,Y) = -\nabla^2\mdelta(X-Y) \times \frac{2C}{\beta^2} \frac{D_{\mathrm{e}}}{L^2}.
\end{equation}
We emphasize that (\ref{eq:stochastichydro}) is also valid beyond linear response and determines the noise strength for arbitrarily large fluctuations in $E$.

It has recently been emphasized in \cite{rangamani, liu15, jensen} that the formal effective field theory approach to hydrodynamics can describe fluids with non-Gaussian and non-Markovian noise spectra.   Our approach implies that non-Gaussian noise is not generic in the long wavelength limit and is suppressed by powers of $L$.   This suggests that in the effective action formalism for hydrodynamics, most nonlinearities involving noise are irrelevant under renormalization group flow.

\section{Analysis of the Fokker-Planck Equation}\label{sec:longhydro}
We have already seen that (\ref{eq:stochastichydro}) is equivalent to (\ref{eq:FPE}). Our analysis explains when and how hydrodynamics arises in a quantum many-body system.    The purpose of this section is to clarify a number of further features of (\ref{eq:stochastichydro}) which are \emph{guaranteed} in generic quantum systems.   In particular, we will elucidate the microscopic origins of the factorizability of near-equilibrium correlation functions and explain why local operators such as $H(X)^2$, $H(X)^3$, etc. are not additional slow degrees of freedom.   Along the way, we will observe an analogy between perturbatively nonlinear hydrodynamics and the Boltzmann theory of weakly interacting gases.

\subsection{Hermite Basis}
We begin by describing how to solve (\ref{eq:FPE}) without invoking stochastic calculus.    Since (\ref{eq:FPE}) is linear, it is natural to look for a convenient basis of functions in which to write the solution.   We will focus our discussion by thinking about the dynamics close to global thermal equilibrium at temperature $T=1/\beta$.   Observe that the canonical ensemble \begin{equation}
p_{\mathrm{eq}}(\mathbf{E}) = \prod_X \frac{\omega(E(X)) \mathrm{e}^{-\beta E(X)}}{Z(\beta)}  \label{eq:canonical}
\end{equation}
with $Z(\beta)$ the partition function in any given region, is an exact solution to (\ref{eq:markov}) and (\ref{eq:FPE}).    To see this explicitly, observe that \begin{equation}
\sum_{Y} D \Lambda(X,Y) \partial_{E(Y)} \left(\frac{p_{\mathrm{eq}}}{\Omega}\right) \propto \sum_{Y:Y\sim X} D(E(X),E(Y)) \left(\partial_{E(X)} - \partial_{E(Y)}\right) \exp\left[-\beta\sum_{X^\prime}E(X^\prime )\right] = 0.
\end{equation}

(\ref{eq:canonical}) forms the starting point for a convenient basis set.    First, we observe that $\log\omega(E) \propto L^d$ is an exponentially large function when $E\propto L^d$ is extensive;  this is the regime of interest as it corresponds to finite temperature.   Therefore, as is well known from statistical physics: 
\begin{equation}
\frac{\omega(E)\mathrm{e}^{-\beta E}}{Z(\beta)} \approx \frac{1}{\sqrt{2\mpi}\sigma}\exp\left[-\frac{(E-\bar E(\beta))^2}{2\sigma^2}\right] \label{eq:gaussian}
\end{equation}
where \begin{subequations}\begin{align}
\bar E(\beta) &= - \frac{\partial \log Z}{\partial \beta}, \\
\sigma^2 &=  -\frac{\partial \bar E}{\partial \beta}
\end{align}\end{subequations}
Note that $\bar E \sim \sigma^2 \sim L^d$, and that $\sigma^2 = T^2 C$ is proportional to the specific heat $C$ of each region.   Energy fluctuations are extremely small in the large $L$ limit.   For future convenience we  define  \begin{equation}
\eta = \frac{E-\bar E}{\sigma}.
\end{equation}
We define the Hermite polynomials for non-negative integers $n$: \begin{equation}
\mathrm{H}_n(x) = \frac{1}{\sqrt{n!}} \mathrm{e}^{x^2/2} \left(-\frac{\mathrm{d}}{\mathrm{d}x}\right)^n \mathrm{e}^{-x^2/2}, \label{eq:hermite}
\end{equation}
which form an orthonormal basis of polynomials: 
\begin{equation}
\int \mathrm{d}x \; \mathrm{H}_n(x)\mathrm{H}_m(x) \frac{\mathrm{e}^{-x^2/2}}{\sqrt{2\mpi}} = \mdelta_{nm}.
\end{equation}
Note that our normalizations are different from those in standard physics textbooks.     For future convenience, we note the following useful identities: \begin{subequations}\label{eq:hermiteidentities}\begin{align}
\partial_x \mathrm{H}_n(x) &= \sqrt{n}\mathrm{H}_{n-1}(x), \\
x\mathrm{H}_n(x)  &=  \sqrt{n+1}\mathrm{H}_{n+1}(x) + \sqrt{n}\mathrm{H}_{n-1}(x).
\end{align}\end{subequations}

Using these definitions and identities, we now expand $p(\mathbf{E},t)$ in the following basis:  
\begin{equation}
p(\mathbf{E},t) = p_{\mathrm{eq}}(\mathbf{E}) \sum_{\mathbf{n}} c(\mathbf{n},t) \prod_X \mathrm{H}_{n(X)}\left(\eta(X)\right).
\end{equation}
Without loss of generality we may ignore the dynamics of the $\mathbf{0}$ mode, which we have already seen is a steady state solution.   It is convenient to view the probability distribution as a member of a linear space of functionals, with a generic vector denoted $| {\bf f} ]$. Define the inner product \begin{equation}
[ {\bf f} |{\bf f^\prime}] = \int \mathrm{d}\mathbf{E}\; p_{\mathrm{eq}}(\mathbf{E}) f(\mathbf{E}) f^\prime(\mathbf{E}),
\end{equation}
so that \begin{equation}
[ \mathbf{n}|\mathbf{m}] = \prod_X \mdelta_{n(X),m(X)}.
\end{equation}
Defining \begin{equation}
K(\mathbf{n},\mathbf{m}) = \sum_{X\sim Y} \left[ \mathbf{n}\left| \overleftarrow{\partial_{E(X)}} \mathcal{D}(\mathbf{E},X,Y)  \overrightarrow{\partial_{E(Y)}} \right|\mathbf{m}\right], \label{eq:Kdef}
\end{equation}
where $\mathcal{D}$ simply represents multiplication by the function $\mathcal{D}$ (defined in (\ref{eq:calD})), we conclude that so long as we do not reach values of $n$ large enough for the Gaussian approximation (\ref{eq:gaussian}) to fail, the vector \begin{equation}
|\mathbf{p}(t)] = \sum_{\mathbf{n}} c(\mathbf{n},t)|\mathbf{n}]
\end{equation}
evolves in time as \begin{equation}
\partial_t |\mathbf{p}(t)] = - K|\mathbf{p}(t)].    \label{eq:KEOM}
\end{equation}
We will discuss the regime of validity of this ``Gaussian" approximation  in Section \ref{sec:nonlinear}.

\subsection{Diffusion Near Equilibrium}
To proceed further, let us analyze (\ref{eq:KEOM}) carefully when the Gaussian approximation holds.   Since $E(X)\approx \bar E(\beta)$ for every $X$, we can approximate \begin{equation}
\mathcal{D}(\mathbf{E},X,Y) \approx \mathcal{D}_0 \Lambda(X,Y)  \label{eq:D0approx}
\end{equation}
in (\ref{eq:Kdef}); here $\mathcal{D}_0$ is a $\beta$-dependent constant.  (We will relax this assumption later in Section \ref{sec:nonlinear}.)   Using (\ref{eq:Laplacian}), (\ref{eq:hermiteidentities}), (\ref{eq:Kdef}) and (\ref{eq:D0approx}), we find \begin{align}
K(\mathbf{n},\mathbf{m}) &= \frac{\mathcal{D}_0}{\sigma^2} \sum_{X\sim Y} \left( \prod_{Z\ne X,Y} \mdelta_{n(Z),m(Z)} \right)\left( (n(X)+n(Y))\mdelta_{n(X),m(X)}\mdelta_{n(Y),m(Y)} \right. \notag \\ 
&\;\;\;\;\; \left. - \sqrt{m(Y)n(X)} \mdelta_{n(X),m(X)+1}\mdelta_{n(Y),m(Y)-1} - \sqrt{n(Y)m(X)} \mdelta_{n(X),m(X)-1}\mdelta_{n(Y),m(Y)+1}  \right)   \label{eq:DHermite}
\end{align}
The key observation is that $K(\mathbf{n},\mathbf{m})$ is \emph{block diagonal}.   Defining \begin{equation}
|\mathbf{n}| = \sum_X n(X),
\end{equation}
we see that $K(\mathbf{n},\mathbf{m}) \ne 0$ if and only if $|\mathbf{n}| = |\mathbf{m}|$.

The origin of this tower of ``conservation laws" is the fact that (\ref{eq:markov}) admits steady state solutions for arbitrary superpositions of microcanonical ensembles.   Schematically, the conservation of $|\mathbf{n}|$ in the dynamics simply follows from the fact that $\mathcal{F}$, defined in (\ref{eq:calF}), is arbitrary.   Each sector of $|\mathbf{n}|$ probes an orthogonal component to the function $\mathcal{F}$.  Thus, each of these conservation laws is the conservation of energy, but on a different superposition of energy shells.

Let us first analyze $|\mathbf{n}| = 1$.   Here it is natural to refer to basis vectors $|\mathbf{n}\rangle$ by simply the site $X$ on which $n(X)=1$:  $|X]$.   From (\ref{eq:DHermite}) we conclude that $K(\mathbf{n},\mathbf{m})$  (which we will write as $K(X,Y)$ in the $|\mathbf{n}|=1$ sector),  can simply be written as \begin{equation}
K(X,Y) = \frac{\mathcal{D}_0}{\sigma^2}\Lambda(X,Y).  \label{eq:42K0}
\end{equation}
Note that $\mathcal{D}_0 \sigma^{-2} \sim L^{-2}$.   

Considering for the moment a system on a simple one dimensional lattice with periodic boundary conditions, the eigenvalues of $\Lambda(X,Y)$ are given by $2(1-\cos(k L))$ where $k=n \mpi /L$ with $n\in \mathbb{Z}$.    So the spectrum of $K(X,Y)$ is given by \begin{equation}
K \sim \frac{\mathcal{D}_0 L^2}{\sigma^2} k^2  \label{eq:42K}
\end{equation}
at small $k$.   This is precisely the hydrodynamic limit.   The coefficient of $k^2$ in the eigenvalue spectrum is independent of $L$ in the large $L$ limit, and corresponds to the physical energy diffusion constant (\ref{eq:Denergy}).   As we argued previously, (\ref{eq:Denergy}) will also hold on a more generic lattice graph in any spatial dimension whenever the graph is a suitable triangulation of flat space.    In the case of anisotropic lattice models, the low-lying spectrum of $\Lambda(X,Y)$ will exhibit the proper anisotropy and the diffusion constant $D_{\mathrm{e}}$ will become a second-rank tensor.

To justify that (\ref{eq:Denergy}) is the physical energy diffusion constant in the Hermite basis, consider the system whose state is described by the vector $|{\bf p}]$ consisting of only $c(\mathbf{0},t)=1$, along with $|\mathbf{n}|=1$ modes.  From (\ref{eq:42K0}), \begin{equation}
\partial_t c(X,t) = -\sum_Y K(X,Y)c(Y,t).  \label{eq:42K2}
\end{equation}
The probability of measuring energy $E(Y)$ when the system is in  $|\mathbf{p}(t)]$ is given by 
 \begin{equation}
\int \mathrm{d}\mathbf{E} \; E(Y)p(\mathbf{E},t) = [ \mathbf{0}|E(Y) |\mathbf{p}(t)] = \bar E(\beta) + \sigma [ Y|\mathbf{p}(t)] = \bar E(\beta) + \sigma c(Y,t).
\end{equation}
Up to the overall factor of $\sigma$, $c(Y)$ is proportional to $\langle E(Y)\rangle$ and therefore $\langle E(Y,t)\rangle$ also obeys (\ref{eq:42K2}).   Hence $D_{\mathrm{e}}$ is the physical energy diffusion constant.   Reverting to the definition of the energy density as a quantum expectation value, \begin{equation}
\partial_t \langle H(X)\rangle_{\mathrm{ne}} = D_{\mathrm{e}} \nabla^2_X \langle H(X)\rangle_{\mathrm{ne}}  + \cdots  \label{eq:quantumdifflin}
\end{equation}
where $\cdots$ denotes higher derivative, $L$-dependent corrections to diffusion, and $\langle \cdots \rangle_{\mathrm{ne}}$ denotes expectation values of operators in a non-equilibrium initial state obeying $\mathcal{N}\rho(0)=0$.

\subsection{Higher Derivative Corrections}\label{sec:higherderivative}

Although we do recover a sensible diffusive limit independent of the cutoff choice $L$, at higher orders in wave number $k$, the eigenvalue spectrum of $K(X,Y)$ will generally depend on $L$.   This is not surprising, since by construction our approach only contains modes at wave numbers $k\lesssim L^{-1}$.   If $L$ is chosen sufficiently large, then our approach must underestimate the decay times of hydrodynamic, diffusive modes.   A simple model to quantitatively understand this effect consists of starting with our 1d chain model and ``integrating out" every other site.   In this $|\mathbf{n}|=1$ sector, we can perform this integrating out by writing basis vectors $\mathbf{e}(X)$ ($X\in \mathbb{Z}$) as \begin{equation}
|\tilde X^\pm]  = \frac{|2X]  \pm |2X+1]}{\sqrt{2}} .
\end{equation}
(The factor of $\sqrt{2}$ will account for the $\sigma$-dependence of (\ref{eq:42K0})).   Defining the matrices \begin{equation}
\tilde A^\pm(X,Y) = \left\lbrace \begin{array}{ll} 1 &\ X=Y+1 \\ \pm 1 &\ X=Y-1 \\ 0 &\ \text{otherwise} \end{array}\right.,
\end{equation}
we may write a $2\times 2$ block matrix form for the effective Laplacian:  \begin{equation}
\tilde \Lambda(X,Y) =  \frac{1}{2} \left(\begin{array}{cc} 2 - A^+(X,Y) &\  A^-(X,Y) \\ -A^-(X,Y) &\ 6 + A^+(X,Y)  \end{array}\right).  \label{eq:tildeLambda}
\end{equation}
The top row corresponds to $+$ modes; the bottom row to $-$ modes.  Integrating out the $-$ modes, we naively obtain a new dispersion relation by simply truncating $\tilde\Lambda(X,Y)$ to the top left column.   As $2-A^+(X,Y) = \Lambda(X,Y)$ for the 1d lattice, we appear to obtain precisely the same result as prior but with the overall cofficient in the diffusion constant reduced by a factor of 2.    However, this is not correct -- the prefactor should be reduced by a factor of 4.   We cannot neglect the off-diagonal components in (\ref{eq:tildeLambda}), which couple degrees of freedom that we have integrated out and those we have kept.   To estimate the magnitude of this effect, we use the memory matrix formalism again.   The modes that we are integrating out always have a larger decay rate than the modes that we keep (in the absence of off-diagonal coupling).  To leading order in $k$, we may approximate the memory function as time-independent (Markovian), at the cost of finding that \begin{equation}
\tilde\Lambda^{\mathrm{eff}}(X,Y) = \frac{1}{2}\left(\Lambda + \frac{1}{8} \left(A^-\right)^2   \right)(X,Y).
\end{equation} 
In the long wavelength limit, the eigenvalues of $\tilde\Lambda^{\mathrm{eff}}(X,Y)$ are of the form $\mathrm{e}^{\mathrm{i}kX}$, with $\tilde\Lambda^{\mathrm{eff}}(k) \approx \frac{1}{4}k^2$.  Since $\Lambda(k) \approx k^2$ as $k\rightarrow 0$, we correctly recover the diffusive limit with the right relative scaling.

There are two points worth noting.   Firstly, the matrix structure of $\tilde\Lambda^{\mathrm{eff}}$ differs from nearest neighbor in this simple model.   Secondly, at higher orders in derivatives, corrections to the Markovian assumption become necessary, even though the decay rate of the $+$ mode is still smaller than the $-$ mode.   These facts can all be taken into account in the memory function formalism.  The key approximation which would have to be relaxed is that $\mathcal{NLN} \approx \sum_X \mathcal{L}(X)$.   As our focus here is on the universal hydrodynamic limit -- the physics for $k\sim L^{-1}$ is not universal and is sensitive to the cut-off -- we will not worry about these subtleties with the Markovian, nearest-neighbor approximations for the remainder of the paper.

\subsection{Do Non-hydrodynamic Slow Modes Exist?}
Next we turn to the $|\mathbf{n}|=2$ solutions to the near equilibrium FFPE.    Here a useful basis choice is $|XY]$, which denotes the vector where $n(X)=n(Y)=1$ (if $X\ne Y$), and $n(X)=2$ (if $X=Y$).   $|XY] = |YX]$ are equivalent.  We emphasize that mode $|XY]$ is orthogonal to any of the energy fluctuations $|X]$ in the $|\mathbf{n}|=1$ sector; an observer measuring local Hamiltonians $H(X)$ does not access fluctuations in the $|\mathbf{n}|=2$ sector.    Instead, what $|XY]$ corresponds to is a (possibly long-ranged) correlation between two distinct fluctuations:  \begin{equation}
[XY|\mathbf{p}] \propto \langle (H(X) - \bar E) (H(Y) - \bar E)\rangle_{\mathrm{ne}}.  \label{eq:HXYcumulant}
\end{equation} 

\begin{figure}[t]
\centering
\includegraphics[width=3.3in]{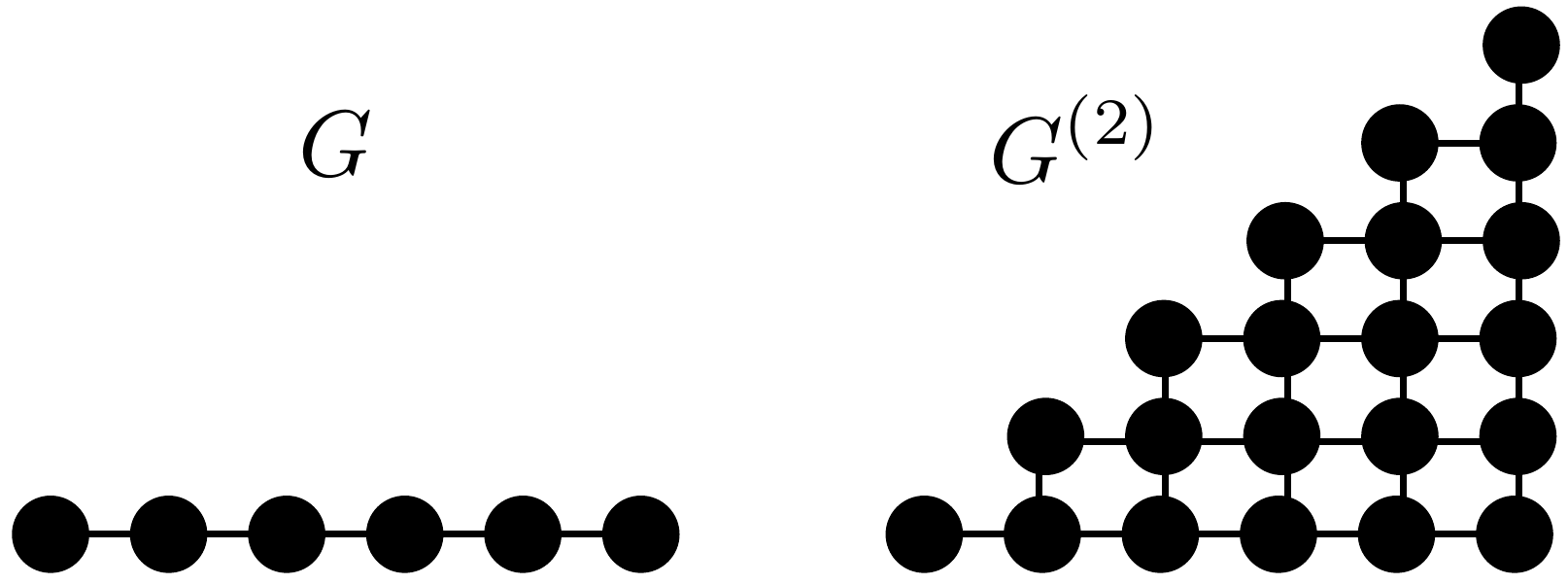}
\caption{At higher $|\mathbf{n}|$, the FFPE reduces to a discrete Markovian diffusion problem on a higher dimensional analogue of the graph $G$, as described in (\ref{eq:G2}).   Except near the diagonal boundary in $G^{(2)}$ above, the transition rates for the Markov process are separable in each of $|\mathbf{n}|$ ``dimensions."   The physics of the $|\mathbf{n}|=1$ sector thus largely controls higher $|\mathbf{n}|$ sectors as well (for finite $|\mathbf{n}|$). }
\label{fig:simplex}
\end{figure}

Let us first describe the transitions between these modes when $\mathrm{dist}(Y,X)>1$; here dist simply denotes the distance between the two sites on graph $G$.     Because (\ref{eq:DHermite}) describes the hopping of excitations between adjacent sites on the superlattice $G$, when $\mathrm{dist}(Y,X)>1$ the transition matrix $K$ can be thought of a sum of two transition matrices, one for each `excitation'.   In other words, \begin{equation}
K(XY,X^\prime Y^\prime) = K(X,X^\prime) \otimes \mdelta(Y,Y^\prime)  + \mdelta (X,X^\prime) \otimes K(Y,Y^\prime), \;\;\;\;\; (\mathrm{dist}(Y,X),\mathrm{dist}(Y^\prime,X^\prime) > 1).  \label{eq:KG2}
\end{equation}
Transition rates betwen $|XX]$ and $|XY]$ in (\ref{eq:DHermite}) contain additional factors of $\sqrt{2}$ and 2, but this can be removed  by simply rescaling the $|XX]$ basis vector by a factor of $\sqrt{2}$.   In the rescaled basis, $K(XY,X^\prime Y^\prime)$ is proportional to a graph Laplacian $\Lambda^{(2)}(XY,X^\prime Y^\prime)$ on a graph $G^{(2)} = (V^{(2)}, \mathcal{E}^{(2)})$ with \begin{subequations}\label{eq:G2}\begin{align}
V^{(m)} &= V^m/\mathrm{S}_m, \\
\mathcal{E}^{(m)} &= \left(\bigcup_{j=1}^m \bigcup_{e=(u_jw_j)\in\mathcal{E}, u_i\in V} \lbrace (u_1\cdots u_m,u_1\cdots u_{j-1}w_j u_{j+1}\cdots u_m) \rbrace \right) / \mathrm{S}_m.
\end{align}\end{subequations}
In the above equation we take $m=2$ for now.  The group action $\mathrm{S}_2$ swaps the two vertices between the copies of $V$; we identify the orbit of each element under the group action as a unique element in either $V^{(2)}$ or $\mathcal{E}^{(2)}$.   For example, if $G=(V,\mathcal{E})$ corresponds to a one dimensional lattice of $N$ sites, then $G^{(2)}$ is a two-dimensional simplex of $N(N+1)/2$ sites:  see Figure \ref{fig:simplex}.   

Because (\ref{eq:KG2}) is (away from the $XX$ sites) a \emph{separable} transition matrix, we can estimate the eigenvalue spectrum from the eigenvalue spectrum of $K(X,X^\prime)$.   Since the eigenvalue spectrum of $K(X,X^\prime)$ is diffusive on long length scales, for a quantum system in $d$ dimensions, there are $N(\omega)\sim \omega^{d/2}$ eigenvalues $\le \omega$.    The number of eigenvectors of $K(XY,X^\prime Y^\prime)$ can then be estimated as \begin{equation}
N_m(\omega) \approx \frac{N(\omega)^m}{m!} \label{eq:Nmomega}
\end{equation}
with $m=2$.   Justification for this claim follows from plugging $\mathrm{S}_2$-invariant sums of tensor products of eigenvectors into the variational principle applied to the real symmetric matrix $K$ -- the vectors $|XX]$ lead to $\mathrm{O}(N^{-1})$ corrections.    Thus the vast majority of $G^{(2)}$ is described by separable dynamics where the two excitations effectively undergo independent, simultaneous random walks.   

In more physical terms, what we have found is that the correlators of the energy density \emph{factorize}.   This is a well-known property of finite temperature correlation functions.   We emphasize that in our formalism we did \emph{not} assume  this property upfront.   Instead, factorization is a consequence of our earlier Markovian assumption (\ref{eq:markov}).    To be explicit, in the continuum limit, (\ref{eq:quantumdifflin}) generalizes to \begin{align}
\partial_t \langle H(X,t)H(Y,t)\rangle_{\mathrm{ne}}= D_{\mathrm{e}} \left(\nabla_X^2 + \nabla_Y^2\right)  \langle H(X,t)H(Y,t)\rangle_{\mathrm{ne}}+ \cdots   \label{eq:quantumdifflin2}
\end{align}
Using standard linear response theory, a common way of weakly perturbing thermal equilibrium to obtain a non-equilibrium state is simply to act with an operator such as $H(X^\prime)H(Y^\prime)$ on the thermal density matrix at time $t=0$.   For such a non-equilibrium initial condition (\ref{eq:quantumdifflin2}) will explicitly factorize, so long as $X$ and $Y$ are reasonably well separated:  \begin{equation}
\partial_t \langle H(X,t)H(Y,t) H(X^\prime,0)H(Y^\prime,0) \rangle_\beta  = D_{\mathrm{e}} \left(\nabla_X^2 + \nabla_Y^2\right)  \langle H(X,t)H(Y,t) H(X^\prime,0)H(Y^\prime,0)\rangle_\beta+ \cdots.
\end{equation}
Here $\langle \cdots\rangle_\beta$ denotes the standard thermal average.  The above identities are a fingerprint of hydrodynamics in a thermalizing quantum many-body system. (\ref{eq:quantumdifflin2}) makes clear that the slow degrees of freedom in the $|\mathbf{n}|=2$ sector simply correspond to two hydrodynamic excitations which have almost completely decoupled.  Hence the slow dynamics in the $|\mathbf{n}|=2$ sector should not be thought of as new physical degrees of freedom, even though they probe a part of the probability distribution $|{\bf p}]$ that is invisible to the energy fluctuations in the $|\mathbf{n}|=1$ sector.    Also observe that at late times, to very good approximation (\ref{eq:quantumdifflin2}) is true even if the diffusive ``clouds" overlap (i.e. $[XX|\mathbf{p}] \ne 0$).   So the factorization (\ref{eq:quantumdifflin2}), which follows directly from the Markovian assumption (and locality), is slightly stronger than simply an assumption that thermal correlation functions factorize on large length scales.
 
From a mathematical perspective, the separability of $K$ along with the trivial eigenvalue spectrum (\ref{eq:Nmomega}) is analogous to the mathematical origin of the truncation of the Bogoliubov-Born-Green-Kirkwood-Yvon (BBGKY) hierarchy in kinetic theory which reduces the Liouville equation for $N$ particles to the Boltzmann equation for a single particle.   While a generic $N$-particle distribution function evolves in a very complicated manner, the physical content of the BBGKY hierarchy is that a special initial condition in which cumulants of simple quantities vanish will not lead to the development of large cumulants on late time or length scales \cite{pulvirenti}.   In our simplified limit of near-equilibrium fluctuations, this is precisely guaranteed by the decoupling of sectors at different $|\mathbf{n}|$.  If all cumulants (\ref{eq:HXYcumulant}) vanish at $t=0$, they will vanish for all times $t$.   As in kinetic theory, the truncation of the hierarchy is rather trivial at this point, but will become less so when we relax (\ref{eq:D0approx}) in  Section \ref{sec:nonlinear}.   We also strongly emphasize the conceptual differences between kinetic theory and our formalism.  In kinetic theory, the factorization analogous to (\ref{eq:quantumdifflin2}) arises from the integrability of non-interacting Hamiltonians;  in our approach this factorizability is instead derived from the chaotic microscopic dynamics and the emergence of dissipation, as described in Section \ref{sec:diss}.  

A final interesting point about the dynamics in the $|\mathbf{n}|=2$ sector is that an observer who can only measure the operators $\langle H(X,t)^2 \rangle $ will see an `intermediate' dynamics which is neither hydrodynamic nor decaying exponentially to its steady state value.  In the $|\mathbf{n}|=2$ sector, such measurements amount to the weight of the distribution on $|XX]$ sites, which will decay algebraically with time as the weight diffuses into the ``bulk" of the graph $G^{(2)}$.     Since $H(X,t)^2$ is simply the product of two hydrodynamic operators, the origin of this algebraic decay is intimately related to the hydrodynamic character of the slow dynamics in the $|\mathbf{n}|=2$ sector.   In fact, the algebraic decay in $\langle H(X,t)^2 \rangle_{\mathrm{ne}}$ can be thought of as an avatar of long-time tails \cite{alder}, which arise when non-hydrodynamic operators nonlinearly couple to hydrodynamic operators and therefore decay algebraically instead of exponentially.   As we will see in Section \ref{sec:nonlinear}, the origin of perturbative nonlinearity in the diffusion equation is the coupling of sectors at different $|\mathbf{n}|$, which has so far been neglected due to the approximation (\ref{eq:D0approx}).

Finally, we turn to the case of $|\mathbf{n}|=m$.  As the calculation proceeds analogously to the case $|\mathbf{n}|=2$ we simply summarize the result.  When $m\ll \sqrt{N}$, nearly every point at fixed $|\mathbf{n}|=m$ has $m$ distinct values, as can be seen from the simple combinatoric estimate \begin{equation}
\dfrac{\displaystyle \left(\begin{array}{c} N \\ m \end{array}\right)}{\displaystyle \left(\begin{array}{c} N+m-1 \\ m \end{array}\right)} \approx \exp\left[-\frac{2m^2}{N}\right].
\end{equation} 
(The numerator corresponds to the number of points with $m$ distinct vertices; the denominator is the number of vertices in $V^{(m)}$.)   Whenever this ratio is very close to 1, the late time dynamics in the $m$ sector is (away from the boundary) generated by the graph Laplacian on the graph $G^{(m)} = (V^{(m)}, \mathcal{E}^{(m)})$ defined in (\ref{eq:G2}).   The steady state solution is then uniform (at least away from the boundary).   We are overwhelmingly likely to find that a single excitation is present on $m$ distinct sites.    Following the logic above, we conclude that for all such states, the $m$-point generalization of (\ref{eq:quantumdifflin2}) holds.   Although the number of slow degrees of freedom is given by the extremely large number (\ref{eq:Nmomega}), these slow degrees of freedom do not correspond to physically distinct modes -- they are (mostly) decoupled hydrodynamic excitations evolving independently.   Therefore there are no slow degrees of freedom beyond the hydrodynamic modes.
 
\subsection{Coherent Basis}
When $|\mathbf{n}| \sim N$, the analysis in a sector of fixed $|\mathbf{n}|$ is no longer convenient.   Just as we passed from the microcanonical ensemble to the canonical ensemble to evaluate the FFPE, it is natural to pass to a ``canonical ensemble" for the $\mathbf{n}$-basis.  In other words, we will study a basis of ``coherent states" (named by their obvious analogy to coherent states of the quantum harmonic oscillator) in our linear space of probability functionals:
 \begin{equation}
|\mathbf{a} ] =  \sum_{\mathbf{n}} \prod_X \frac{a(X)^{n(X)}}{\sqrt{n(X)!}} |\mathbf{n}] 
\end{equation}
Note that since $[\mathbf{0}|\mathbf{a}] = 1$, every single coherent state describes a normalized probability distribution.    These coherent states admit a very simple interpretation:  using (\ref{eq:gaussian}) and (\ref{eq:hermite}) we obtain that $|\mathbf{a}]$ corresponds to the probability distribution $p_{\mathbf{a}}$ given by \begin{equation}
p_{\mathbf{a}} = \prod_X \sum_{n(X)=0}^\infty \frac{(-a(X)\partial_{\eta(X)})^n}{n!} \frac{\mathrm{e}^{-\eta(X)^2/2}}{\sqrt{2\mpi}\sigma}= \prod_X \frac{\mathrm{e}^{-(\eta(X)-a(X))^2/2}}{\sqrt{2\mpi}\sigma}.
\end{equation}
When $a(X)$ is sufficiently small, we can interpret $a(X)$ as a shift in the local inverse temperature: \begin{equation}
a(X) = \frac{\langle H(X)\rangle - \bar E(\beta)}{\sigma}.
\end{equation}
%   Since  \begin{equation}
%p_{\mathbf{a}}(\mathbf{E}) =  \prod_X \frac{\mathrm{e}^{-(E(X)-\bar E(\beta) - \sigma a(X))^2/2\sigma^2}}{\sqrt{2\mpi}\sigma}
%\end{equation}
The regime $a(X)\sim 1$ for every $X$ is the regime in which the model of independent hydrodynamic excitations fails.  

These coherent states are elegant because even if $a(X)\gg 1$, their time evolution is still simple.  To show this explicitly, we consider a state which is a superposition of coherent states: \begin{equation}
|\mathbf{p}(t)] = \int \mathrm{d}\mathbf{a} \;     P(\mathbf{a},t) |\mathbf{a}].  \label{eq:staycoherent}
\end{equation}
Note that the requirement that $1=[\mathbf{0}|p(0)]$ implies that $P(\mathbf{a},0)$ can be interpreted as a \emph{new probability distribution} over the space of all $\mathbf{a}$.   Because $|\mathbf{a}]$ is an overcomplete basis set, we may write (\ref{eq:staycoherent}) for all $t$ and replace the evolution equation (\ref{eq:KEOM}) for $|\mathbf{p}(t)]$ with a new equation of motion for $P(\mathbf{a},t)$.    To compute $\partial_t P(\mathbf{a},t)$, note the following identities for a single box $X$: \begin{subequations}\label{eq:coherentidentities}\begin{align}
\eta |a] &= (a+\partial_a)|a], \\
\overleftarrow{\partial_\eta} |a] &= a|a], \\
\overrightarrow{\partial_\eta}|a] &= \partial_a |a].
\end{align}\end{subequations} 
These identities may be derived by left multiplying by $[{\bf n}|$ and observing that the set of $n(X)$ form a complete basis. Using (\ref{eq:Kdef}) we thus find  
\begin{equation}
\partial_t |p] = - \int \mathrm{d}\mathbf{a} \; P(\mathbf{a})  K | \mathbf{a}] = \int \mathrm{d}\mathbf{a} \;   \sum_{X\sim Y} \frac{\mathcal{D}_0}{\sigma^2} (\partial_{a(X)} - \partial_{a(Y)}) \left((a(X)-a(Y)) P(\mathbf{a})\right)    |\mathbf{a}]
\end{equation}
In the second equation above, we have integrated by parts.    This suggests that we interpret \begin{equation}
\partial_t P = \sum_{X\sim Y} \frac{\mathcal{D}_0}{\sigma^2} (\partial_{a(X)} - \partial_{a(Y)}) \left((a(X)-a(Y)) P\right) = \sum_{X,Y} K(X,Y) \partial_{a(X)} (a(Y)P),
\end{equation}
which we recognize as a simple transport equation.  Whenever $P(\mathbf{a},0) = \mdelta(\mathbf{a}-\mathbf{a}(0))$:
\begin{equation}
\partial_t a(X) = -\sum_{X,Y} K(X,Y) a(Y).  \label{eq:detaeq}
\end{equation}

A few more comments on (\ref{eq:detaeq}) are in order.   Firstly, near equilibrium at fixed inverse temperature $\beta$,  the deterministic equation (\ref{eq:detaeq}) is \emph{equivalent} to the linearized stochastic equation (\ref{eq:stochastichydro}) for certain Gaussian initial conditions where $P(\mathbf{a},0) = \mdelta(\mathbf{a}-\mathbf{a}(0))$.    This is analogous to the entirely classical motion of coherent states of the quantum harmonic oscillator.   In this case, the fluctuations are thermal and not quantum mechanical;  within linear response the fluctuations remain stationary for all time.   For these special initial conditions, in the long wavelength limit, the many-body dynamics of the quantum system has truly reduced to a classical deterministic diffusion equation.

\subsection{Nonlinearity in the Coherent Basis}\label{sec:nonlinear}
The coherent state model, as we have introduced it above, becomes formally inaccurate when $a(X) \sim L^{d/2}$.   In this regime, the effective value of $\beta$ changes by a large O(1) factor, and $\mathcal{D}(\mathbf{E},X,Y)$ along with $\sigma$ can no longer be approximated by their near equilibrium values.     

In order to get some understanding for what happens in this limit, let us first continue to assume that $a(X)$ is sufficiently small so that the width of the Gaussian is unchanged, but relax the assumption that $\mathcal{D}$ can be approximated as (\ref{eq:D0approx}).   We now consider \begin{equation}
\mathcal{D}(\mathbf{E}, X,Y) \approx \mathcal{D}_*\left(\frac{E(X)+E(Y)}{2}-\bar{E}(\beta)\right) \Lambda(X,Y)   \label{eq:calDnonlinear}
\end{equation}
where $\bar{\mathbf{E}}(\beta) = (\bar E(\beta), \cdots, \bar E(\beta))$.   As stated in Section \ref{sec:hydro}, we need not consider gradient corrections to $\mathcal{D}$ because there are other gradient corrections our approach has already missed.  %Furthermore, we will not consider quadratic corrections in $E(X)+E(Y)$ because we are still close to equilibrium.    

We now compute the corrections to $\partial_t P(\mathbf{a},t)$.   Using (\ref{eq:Kdef}) and (\ref{eq:coherentidentities}), we find
\begin{equation}
\overleftarrow{\partial_{\eta(X)}} \mathcal{D}(E - \bar E(\beta)) \overrightarrow{\partial_{\eta(Y)}} |\mathbf{a}] = a(X) \mathcal{D}(\sigma a + \sigma \partial_a,X,Y) \partial_{a(Y)}  |\mathbf{a}]
\end{equation}
Therefore,
\begin{equation}
\partial_t P = \sum_{X,Y} \frac{\Lambda(X,Y)}{\sigma^2} \partial_{a(X)} \left(a(Y) \mathcal{D}\left( \sigma \frac{a(X) + a(Y)}{2} -\sigma \frac{\partial_{a(X)} + \partial_{a(Y)}}{2},X,Y\right) P\right) .  \label{eq:Pnonlin}
\end{equation}
Unlike (\ref{eq:FPE}), this does not resemble a conventional stochastic equation.  For example, expanding $\mathcal{D}$ to just linear order, it is generally not the case that the second derivative contributions have positive-definite weight.  However since the equation is mathematically equivalent (by changing variables to go back to the $\mathbf{E}$-basis) to (\ref{eq:stochastichydro}), it is a stochastic equation in disguise.    

%Despite the lack of a simple stochastic interpretation for (\ref{eq:Pnonlin}), this is simply a stochastic equation in disguise.  To deal with the function of derivatives $\mathcal{D}$, we define \begin{equation}
%Q(\boldsymbol{\mbeta}) =  \int \mathrm{d}\mathbf{a} \; P(\mathbf{a}) \exp\left[-\frac{1}{2}\sum_X (a(X)-b(X))^2\right].
%\end{equation} 
%Using the identity \begin{equation}
%a(X) \exp\left[-\frac{1}{2}\sum_X (a(X)-b(X))^2\right] = (b(X) + \partial_{b(X)}) \exp\left[-\frac{1}{2}\sum_X (a(X)-b(X))^2\right],
%\end{equation}
%along with the analogue (swapping $\mathbf{a}$ and $\boldsymbol{\mbeta}$) we obtain 
%\begin{equation}
%\partial_t Q = \sum_{X,Y} \frac{\Lambda(X,Y)}{\sigma^2} \partial_{b(X)}\left(\mathcal{D}\left(\sigma \frac{b(X)+b(Y)}{2}\right) (b(Y)+\partial_{b(Y)})Q\right).
%\end{equation}
%In fact, this is (\ref{eq:stochastichydro}).  Indeed, we recognize that 

The function $\mathcal{D}(\mathbf{E} - \bar{\mathbf{E}}_\beta,X,Y) = \mathcal{D}(\sigma \mathbf{a},X,Y)$  varies over the scale  $E(X)-\bar E(\beta)\sim L^d$, or $a\sim L^{d/2}$.     Now suppose that we rescale $\mathbf{a} = L^\kappa \tilde{\mathbf{a}}$ with $0<\kappa \le \frac{d}{2}$;  then the derivative terms inside of $\mathcal{D}$ in (\ref{eq:Pnonlin}) are suppressed by additional factors of $L$.  Thus, to leading order in the large $L$ limit, we can neglect derivatives inside of $\mathcal{D}$, so long as $P(\mathbf{a})$ is a non-singular function of $\tilde{\mathbf{a}}$ in the large $L$ limit.   Returning to the $\mathbf{a}$ variables, we conclude that (\ref{eq:Pnonlin}) is well approximated by a transport equation whenever $P$ is a sufficiently smooth function.  The resulting transport equation can be recast as a \emph{non-stochastic} differential equation for $a(X)$: \begin{equation}
\partial_t  a(X) = - \sum_{Y} \mathcal{D}\left(\sigma \frac{a(X)+a(Y)}{2}\right) \Lambda (X,Y)  a(Y) . \label{eq:nonlincoherent}
\end{equation} 
%where $\langle a(X)\rangle$ denotes the average of $a(X)$.   

When $a \gtrsim L^{d/2}$,  the effective temperature change between different regions $X$ can be very large, and $\sigma$ is no longer constant.   A natural generalization of  (\ref{eq:staycoherent}) is  \begin{equation}
p(\mathbf{E},t) = \int \mathrm{d}\boldsymbol{\mbeta} \; P(\boldsymbol{\mbeta},t) \prod_X \frac{\omega(E(X)) \mathrm{e}^{-\beta(X,t)E(X)}}{Z(\beta(X,t))}.
\end{equation}
Whenever $\beta(X,t)$ is a slowly varying function of $X$, and $P(\boldsymbol{\mbeta},t)$ is a smooth function on the scale $b\sim L^{-d/2}$, we generalize (\ref{eq:nonlincoherent}) to \begin{equation}
\partial_t  \beta(X,t) = -\sum_{Y} \mathcal{D}\left(\frac{\bar E(\beta(X)) + \bar E(\beta(Y))}{2}\right) \frac{\beta(X)^2 \Lambda(X,Y)}{C(\beta(X))} \beta(Y) \label{eq:dethydro}
\end{equation}
where $\mathcal{D}(\frac{1}{2}(E(X)+E(Y)))\Lambda(X,Y) = \mathcal{D}(\mathbf{E},X,Y)$.  For simplicity we have simply written the average $\langle \boldsymbol{\mbeta}\rangle$ as $\boldsymbol{\mbeta}$.     The factor of $\beta^2/C(\beta)$ is the generalization of $1/\sigma^2$;  this factor depends on the point $X$ at which we are evaluating $\partial_t \beta$.    In the continuum limit, this becomes \begin{equation}
\partial_t  \beta(X,t) = \nabla \cdot \left(D_{\mathrm{e}}\left(\beta(X,t)\right) \nabla   \beta(X,t) \right)  
\end{equation}
with 
\begin{equation}
D_{\mathrm{e}}(\beta) = \frac{\mathcal{D}(E(\beta))L^2\beta^2}{C(\beta)}
\end{equation}
is the energy diffusion constant defined at the local inverse temperature.   

Let us emphasize once more that the deterministic hydrodynamics (\ref{eq:dethydro}) should only be understood as holding over smooth distributions of $P(\boldsymbol{\mbeta},t)$, where the local temperature is only defined precisely at $\mathrm{O}(L^0)$.   The reason the stochastic effects have been neglected is simply that stochastic fluctuations in $\boldsymbol{\mbeta}$ are only relevant at subleading order in $L$.   Unlike in a theory with multiple conserved charges \cite{huse, kovtun14}, there are no long-time tails or other stochastic corrections to $D_{\mathrm{e}}$ beyond the ``tree level" result.  

\section{Entropy Production}
In this section, we discuss the irreversibile and dissipative dynamics studied previously in the language of entropy production \cite{robertson, zubarev, ikeda, millis}.   We will see that the second law of thermodynamics is natural in our framework.  More importantly, we will discuss the extent to which entropy production is universal in the hydrodynamic limit and confirm that the hydrodynamic prediction for entropy production out of equilibrium is reproduced in our more microscopic formalism.

\subsection{A Thermodynamic Entropy}
The von Neumann entropy of a many-body quantum system is \begin{equation}
S_{\mathrm{vN}} = -\mathrm{tr}\left(\rho\log\rho\right)
\end{equation} 
and is invariant under time evolution.   It was suggested in \cite{millis} that the thermodynamic entropy in a many-body quantum system should be taken to be \begin{equation}
S_{\mathrm{tot}} = - \mathrm{tr}\left(\rho \mathcal{D} \log \rho\right).
\end{equation} 
Taking this definition, we find that \begin{equation}
S_{\mathrm{tot}} = S_{\mathrm{S}}+S_{\mathrm{T}}
\end{equation}
where $S_{\mathrm{S}}$ denotes a classical Shannon entropy, and $S_{\mathrm{T}}$ denotes the conventional thermodynamic entropy: \begin{subequations}\begin{align}
S_{\mathrm{S}} &= - \sum_{\mathbf{E}} \delta^N p(\mathbf{E}) \log p(\mathbf{E})  =  -\int \mathrm{d}\mathbf{E} \; p(\mathbf{E})\log p(\mathbf{E}), \\
S_{\mathrm{T}} &= - \sum_{\mathbf{E}} \delta^N p(\mathbf{E}) \log \Omega (\mathbf{E})  =  \sum_X \int \mathrm{d}\mathbf{E} \; p(\mathbf{E}) S(E(X)).
\end{align}\end{subequations}
Recall the definition of $S$ in (\ref{eq:Omega}).

Since $p(\mathbf{E})$ evolves under a reversible Markov process, the second law of thermodynamics follows \cite{levin}.  We now explicitly demonstrate this, working for simplicity with the FFPE (\ref{eq:FPE}):  \begin{align}
\partial_t S_{\mathrm{tot}} = -\int \mathrm{d}\mathbf{E} \; (\partial_t p) \log \frac{p}{\Omega} = \sum_{X,Y}\int \mathrm{d}\mathbf{E} \; \frac{\Sigma(\mathbf{E},X,Y)\Omega(\mathbf{E})}{p(\mathbf{E})}  \left(\partial_{E(X)}\frac{p(\mathbf{E})}{\Omega(\mathbf{E})}\right) \left(\partial_{E(Y)}\frac{p(\mathbf{E})}{\Omega(\mathbf{E})}\right) \ge 0.
\end{align}
The inequality above is only saturated when \begin{equation}
\frac{p(\mathbf{E})}{\Omega(\mathbf{E})} = \mathcal{F}\left(\sum_X E(X)\right),  \label{eq:micro6}
\end{equation}
as $\Sigma(\mathbf{E},X,Y)$ has a null vector corresponding to global energy conservation.   If (\ref{eq:micro6}) is not obeyed, then $\partial_t S>0$.   Since $S$ is a function of $p$, we have proven that the only stationary states of our Markov process are (\ref{eq:micro6}), as previously advertised.

In our procedure, we have traced out information on length scales shorter than $L$.  Since $\int \mathrm{d}\mathbf{E} \; p(\mathbf{E}) = 1$, and the spectrum of $E(X)$ is $\propto L^d$, we conclude that \begin{equation}
p(\mathbf{E}) \sim \frac{1}{L^{dN}}.
\end{equation}
This implies that \begin{subequations}\begin{align}
S_{\mathrm{S}} &\sim \log L, \\
S_{\mathrm{T}} &\sim L^d.
\end{align}\end{subequations}
Therefore, the Shannon entropy is subleading;  to leading order in the gradient expansion we may exclusively focus on $S_{\mathrm{T}}$.   In $d>1$, $S_{\mathrm{S}}$ is also subleading to quantum entanglement entropy \cite{lucasbook}, which was explicitly ignored in (\ref{eq:DrhoE}), and scales as $S_{\mathrm{E}} \lesssim L^{d-1} \log L$ in a gapless theory.  For the remainder of this section we will approximate $S_{\mathrm{tot}}\approx S_{\mathrm{T}}$ unless otherwise stated.

\subsection{Hydrodynamic Entropy Production}
As we saw in Section \ref{sec:nonlinear}, to leading order in $L$, it is sensible to approximate the nonlinear stochastic equation (\ref{eq:stochastichydro}) with the deterministic equation (\ref{eq:dethydro}), so long as the local inverse temperature $\beta(X,t)$ is understood to only be defined to extensive order $L^0$.    In this regime,  \begin{equation}
S_{\mathrm{T}} = \sum_X \int\mathrm{d}\mathbf{E} \mathrm{d}\boldsymbol{\mbeta} \; S(E(X)) P\left(L^\kappa (\boldsymbol{\mbeta} - \langle \boldsymbol{\mbeta}\rangle )\right) \prod_Y \frac{\omega(Y)\mathrm{e}^{-\beta(Y)Y}}{Z(\beta(Y))} 
\end{equation}where $P\left(L^\kappa (\boldsymbol{\mbeta} - \langle \boldsymbol{\mbeta}\rangle )\right)$ denotes a sufficiently clustered probability distribution around $\langle \boldsymbol{\mbeta}\rangle $.   Performing the integrals over $\mathbf{E}$ and $\boldsymbol{\mbeta}$, we obtain \begin{equation}
S_{\mathrm{T}} = \sum_X S( \beta(X) ) + \cdots,
\end{equation}
where $\cdots$ denotes subleading orders in $L$.   As before, we will drop the angle brackets on $\beta(X)$ for convenience; we also define $S(E(\beta)) = S(\beta)$.   As expected, the entropy is a local thermodynamic function.

The rate of change of the entropy is given by \begin{equation}
\partial_t S_{\mathrm{T}} = \sum_X \partial_\beta S(\beta(X)) \partial_t \beta(X).
\end{equation}
Using the thermodynamic identity \begin{equation}
\partial_T E = C = T\partial_T S,
\end{equation}
with $T=1/\beta$, we find that \begin{equation}
\partial_t S_{\mathrm{T}} = \sum_{X,Y} \mathcal{D}\left(\frac{\beta(X)+\beta(Y)}{2}\right) \Lambda(X,Y)  \beta(X) \beta(Y)  \label{eq:dtST}
\end{equation}
In the continuum limit, (\ref{eq:dtST}) becomes \begin{equation}
\partial_t S_{\mathrm{T}} = \int \mathrm{d}^dX \; L^2 \mathcal{D} (\nabla \beta)^2 = \int \mathrm{d}^dX\; \beta \left(-\nabla \cdot \left(\mathcal{D}\nabla \beta \right)\right) \label{eq:dtST2}
\end{equation}
Entropy production is non-negative and local, in accordance with the local second law of thermodynamics.

Using the first law of thermodynamics \begin{equation}
\mathrm{d}S(X) = \beta(X)\mathrm{d}E(X),
\end{equation}
(\ref{eq:dtST2}) is equivalent to the nonlinear hydrodynamic equation (\ref{eq:diffusioneq}).  The key observation is that $S_{\mathrm{T}}$ was computed microscopically, and is in complete quantitative agreement with the phenomenology of hydrodynamics.  In fact, as in conventional hydrodynamics, we can interpret (\ref{eq:dtST2}) as the statement that an entropy current with non-negative divergence exists.   The entropy current defined along each edge of the superlattice $X\sim Y$ is \begin{equation}
J_{\mathrm{S}}(X\rightarrow Y) = \beta(X) \mathcal{D}\left(\frac{\beta(X)+\beta(Y)}{2}\right) (\beta(X)-\beta(Y)).
\end{equation}
The rate of change of the local entropy can be derived analogously to (\ref{eq:dtST}): \begin{equation}
\partial_t S(b(X,t)) = \sum_{Y\sim X} J_{\mathrm{S}}(X\rightarrow Y).
\end{equation}
The local second law of thermodynamics is the inequality \begin{equation}
J_{\mathrm{S}}(X\rightarrow Y) + J_{\mathrm{S}}(Y\rightarrow X) \ge 0.
\end{equation}
The observable part of the von Neumann entropy, introduced in \cite{millis}, obeys the local second law of thermodynamics.  More importantly, the actual flow of this observable entropy between regions $X$ is exactly equal to the phenomenological entropy current introduced in Landau's formulation of hydrodynamics \cite{landau}.

\section{Outlook}
In this paper we have presented a complete and unambiguous derivation of the nonlinear stochastic hydrodynamics for energy diffusion in a quantum many-body system on a lattice, to first order in the gradient expansion.   Our derivation solves or clarifies a number of longstanding issues in the literature: in particular, we have recovered (not imposed) the fluctuation dissipation theorem, demonstrated the equivalence of a certain microscopic entropy flow to Landau's phenomenological entropy current, and provided an explicit (although abstract) formula for the energy diffusion constant.   

Although we have relied on ``old-fashioned" methods, our formalism may be useful for addressing a number of modern challenges in physics.   First and foremost, as stated in the introduction, it is a completely open question how to compute diffusion constants in lattice models using an unbiased algorithm.    Our approach provides an algorithm.   To be explicit, consider a non-integrable one dimensional lattice model of spin-$\frac{1}{2}$ degrees of freedom.  Dividing up the lattice into regions with $L \lesssim 14$ sites, we may exactly diagonalize $H(X)$ within each region on a modern laptop computer.   Since transitions between eigenstates induced by $H(X,Y)$ only exchange a small amount of energy, we expect it is possible to numerically compute $W(\mathbf{E},\mathbf{E}^\prime)$ with good accuracy.   Varying the value of $L$ allows the numericist to estimate finite size effects.   Keeping in mind the caveats about our approach for one-dimensional models, we hope that our formalism can help solve the longstanding problem of computing diffusion constants in one dimensional lattice models.  Alternatively, our approach may also help to justify recently proposed algorithms \cite{altman, wurtz} to compute diffusion constants in $d=1$.   Even in $d=2$, such an algorithm may prove immensely valuable, especially in the presence of an additional discrete or continuous symmetry.     

Our approach is  ``backwards" from the usual effective field theory.   We have started from microscopic dynamics and argued what the emergent dynamics is.   In particular, we have justified the statement that there are generically no degrees of freedom beyond the expected hydrodynamic ones.   Perhaps our approach will also be useful in providing intuitive understandings for the exotic thermal supersymmetry algebras and ``world volume" approaches to hydrodynamics advocated in \cite{rangamani, liu15, jensen}.

While this paper deals with quantum models on a lattice, there appears to be no \emph{physical} obstruction to studying field theories in the continuum, which will also have a conserved momentum if translation invariant.  How to generalize our formalism to translation invariant continuum field theory is an interesting open question, especially as our division of the Hilbert space into spatial blocks explicitly breaks translation invariance. 

Finally, it may be interesting to compare our formalism to the theory of random unitary circuits (RUCs) \cite{nahum16, nahum, tibor, khemani, tibor2}.   RUCs are equivalent to a  discrete time Markov chain, and could provide a useful check of when our Markov process (\ref{eq:markov}) breaks down in quantum systems.  For example, there are localized and diffusive ``bound states" in a certain one dimensional RUC \cite{shenker1803} which describe short distance physics beyond (\ref{eq:markov}).   Beyond such short distance effects, we anticipate that our approach is consistent with RUC physics, and may also shed light into the possibility of emergent thermodynamics in RUCs with conservation laws.

\addcontentsline{toc}{section}{Acknowledgements}
\section*{Acknowledgements}
We thank Mukund Rangamani and Steve Shenker for helpful discussions.    AL is supported by the Gordon and Betty Moore Foundation's EPiQS Initiative through Grant GBMF4302.

\bibliographystyle{unsrt}
\addcontentsline{toc}{section}{References}
\bibliography{quantumboundbib}

\end{document}